\documentclass{article}

\usepackage{amsmath,amssymb,amsfonts,amsthm}
\usepackage[all]{xy}
\usepackage{color}

\usepackage{rotating}
\usepackage{fullpage}

\def\endofproof {\hfill{$\Box$}\\}

\renewcommand{\(}{\begin{equation}}
\renewcommand{\)}{\end{equation}}
\newcommand{\bea}{\begin{eqnarray}}
\newcommand{\eea}{\end{eqnarray}}

\def\nnu {\boldsymbol{\nu}}

\theoremstyle{plain}
\newtheorem{theorem}{Theorem}[subsection]

\newtheorem{proposition}{Proposition}[subsection]
\newtheorem{corollary}{Corollary}[subsection]

\theoremstyle{definition}
\newtheorem{definition}{Definition}[subsection]
\newtheorem{example}{Example}[subsection]
\newtheorem{examples}{Examples}[subsection]
\newtheorem{remark}{Remark}[subsection]
\newtheorem{observation}{Observation}[subsection]

\begin{document}

\title{
The $E_8$ moduli 3-stack of the C-field in M-theory
}

\author{Domenico Fiorenza, Hisham Sati, Urs Schreiber}
\maketitle

\begin{abstract}

  The higher gauge field in 11-dimensional supergravity -- the C-field --
  is constrained by quantum effects to be a cocycle in some
  twisted version of differential cohomology. We argue that
  it should indeed be a cocycle
  in a certain twisted nonabelian differential cohomology. 
  We give a simple and natural characterization of the full
  smooth moduli 3-stack of configurations of the C-field, the 
  gravitational field/background, and 
  the (auxiliary) $E_8$-field.  
  We show that the truncation of this moduli 3-stack to a bare 1-groupoid
  of field configurations reproduces the differential integral Wu structures that Hopkins-Singer
  had shown to formalize Witten's argument on the nature of the C-field. 
  We give a similarly simple and natural characterization of the moduli
  2-stack of boundary C-field configurations and show that it is equivalent
  to the moduli 2-stack of anomaly free heterotic supergravity field configurations. 
  Finally we show how to naturally encode the Ho{\v r}ava-Witten boundary condition
  on the level of moduli 3-stacks, and refine it from a condition on 3-forms to
  a condition on the corresponding full differential cocycles. 
\end{abstract}

\medskip
\medskip
\medskip

\tableofcontents

\newpage

%%%%%%%%%%%%%%%%%%%%%%%%%%%%%%%%%%%%%%%%%%%%
\section{Introduction}
%%%%%%%%%%%%%%%%%%%%%%%%%%%%%%%%%%%%%%%%%%%%

The higher gauge fields appearing in string theory 
(such as the $B$-field and the RR-fields) and in 11-dimensional
M-theory (the C-field) have local presentations by
higher degree differential forms that generalize the ``vector potential'' 1-form
familiar from ordinary electromagnetism. However, just as Dirac charge quantization
asserts that \emph{globally} the field of electromagnetism is of a more
subtle nature, namely given by a connection on a circle bundle, 
the higher gauge fields in string theory are globally of a more subtle
nature: they are cocycles in \emph{differential cohomology} 
(see for instance \cite{Freed}). 
Moreover, even this refined statement is strictly true only 
when each of these fields is considered in isolation. In the full theory
they all interact with each other and ``twist'' or ``shift'' each other.
As a result, generally the higher gauge fields of string theory are
modeled by cocycles in some notion of \emph{twisted differential cohomology}.
See \cite{HopkinsSinger, Freed, survey} for mathematical 
background and \cite{DFM, BM, SSSIII, FSS, FiorenzaSatiSchreiberII} for applications in this context. 
In this article we discuss the differential cohomology 
of the C-field in 11-dimensional supergravity, twisted by the 
field of gravity in the bulk of spacetime, as well as by the $E_8$-gauge field
on Ho{\v r}ava-Witten boundaries \cite{HoravaWitten} and on M5-branes.

\vspace{3mm}
The general theory of \emph{twisted differential cohomology} 
and its characterization of higher gauge fields in string theory
it to date only partially understood. For instance, it has been 
well established that the underlying bare cohomology that 
controls the interaction of the $B$-field in type II string theory
with the Chan-Paton gauge bundles on D-branes is \emph{twisted K-theory}, and
that for trivial $B$-field the corresponding differential cohomology theory
is \emph{differential K-theory}, but a mathematical construction of 
fully fledged \emph{twisted differential K-theory} has not appeared
yet in the literature (see, however, \cite{CMW, KV}). 
Similarly, partial results apply to the lift of this configuration from
type II to M-theory. It is clear that the C-field \emph{in isolation}
is modeled by cocycles in degree-4 ordinary differential cohomology,
just as the B-field in isolation is modeled by degree-3 differential 
cohomology, and the electromagnetic field by degree-2 differential cohomology.
Less is known about the interaction of the C-field with the degrees of freedom
on branes, which here are M5-branes. In our companion article 
\cite{FiorenzaSatiSchreiberII} we investigated aspects of this interaction. The present
article provides a detailed discussion of the mathematical model of the C-field, as used there.

\vspace{3mm}
The C-field experiences a subtle twist already by its interaction with 
the field of gravity, via the Spin-structure on spacetime.
This was first argued in \cite{WittenFluxQuantization} (we review the argument in 
section \ref{CSWithBackgroundCharge}):
 \emph{the degree-4 integral class $[2G]$ of the 
C-field is constrained to equal the first fractional Pontrjagin class of the
Spin structure modulo the addition of an integral class divisible by 2}.
The interpretation of division by 2 in the flux quantization is 
given in \cite{Sati10} and related to Wu structures in 
\cite{tw,II}.
The flux quantization condition can be viewed as defining 
a twisted String structure \cite{SSSIII}.  
Dependence of the partition function in M-theory 
on the Spin structure is investigated in \cite{Spinc}.
Anomalies of M-theory and string theory on manifolds 
with String structures via 
$E_8$ gauge theory is discussed in \cite{S-String},
and the relation to gerbes is discussed in \cite{S-gerbe}. 
The $\mathbb{Z}_2$-twist of the C-field 
for a fixed background Spin structure has been formalized in 
\cite{HopkinsSinger}, following an argument in 
\cite{Witten96, WittenFluxQuantization}, 
by a kind of twisted abelian differential cohomology (which we review in 
section \ref{DifferentialIntegralWuStructures}).
However, two questions remain:
\begin{enumerate}
  \item 
  On Ho{\v r}ava-Witten boundaries as well as on M5-branes, the C-field interacts
  with \emph{nonabelian} and in fact 
  \emph{higher} nonabelian gauge fields. \emph{What is the proper refinement
  of the corresponding twisted differential cohomology}
  to \emph{non-abelian} differential cohomology?
  
  \item
    More generally, already the field of gravity, in the first-order formulation
	relevant for supergravity, is a cocycle in \emph{nonabelian} differential cohomology
	(a Poincar\'e-connection decomposing into a vielbein and a Spin connection).
	If we do not fix a gravitational background configuration / Spin structure 
	as in the above model: 
	\emph{what is the nonabelian differential cohomology
	that unifies gravity, the C-field and its boundary coupling to $E_8$-gauge fields?}
\end{enumerate}

\vspace{3mm} 
In previous work \cite{SSSI,SSSIII,FSS} we have developed
a more general theory of \emph{nonabelian} differential cohomology
(see \cite{survey} for a comprehensive account),
and have shown that various phenomena in string theory, such as 
Green-Schwarz anomaly cancellation, find their full description
(technically: the full higher moduli stacks 
of field configurations without any background fields held fixed)
in this theory.  Moreover, in \cite{FiorenzaSatiSchreiberII} we have analyzed aspects of the
nonabelian 2-form field on M5-branes using this machinery,
while briefly sketching related aspects of the C-field. Here we 
provide further details.

\vspace{3mm}
We construct and then analyze a model for the C-field in 
nonabelian differential cohomology. We show that it reproduces
the relevant properties of previous models,
mainly \cite{DMW,DFM,FM,Sati10}, and refines them
in the following ways.
\begin{enumerate}
  \item All three gauge fields are dynamical (gravity, C-field, $E_8$-field),
  none is fixed background. In particular, where in previous models the 
  fixed gravitational background is perceived of as a twist of the 
  dynamical C-field, here the twisting is democratic, and in effect
  the whole construction yields a single \emph{twisted differential String structure}
  as introduced in \cite{SSSIII}.
  \item
    As a result, the whole construction is outside the scope of 
	abelian differential cohomology and necessarily lives in higher nonabelian
	differential cohomology. Only truncations and reductions where the
	Spin connection is held fixed and the $E_8$-field is reduced to its
	instanton sector sit in the purely abelian sector, as previously conceived.
		\item 
   The full moduli 3-stack of field configurations is produced by a simple
   and natural homotopy pullback construction. This means
   that not only the gauge transformations, but also their gauge-of-gauge transformations
   as well as their higher gauge transformations, are accounted for. Moreover, the
   \emph{smooth structure} on all this is retained. In summary, this means that 
   the smooth moduli 3-stack that we produce integrates the relevant
   (off-shell) BRST Lie 3-algebroid of field configurations (gravity, C-field, $E_8$-field), 
   involving the appropriate ghosts, 
   ghosts-of-ghosts and third order ghosts. This is the correct starting point for
   any actual quantization of the system (as an effective low-energy 
   gravitational higher gauge theory,
   as it were, but conceivably of relevance also to the full ``M-theory'').
  \item
    A similarly simple and natural further homotopy pullback gives the boundary
	field moduli 2-stack of the C-field. We demonstrate that this is
	equivalent to the moduli 2-stack of anomaly free heterotic field
	configurations as found in \cite{SSSIII}.
  \item
    We lift the Ho{\v r}ava-Witten boundary condition on the C-Field 
	from 3-forms to differential cocycles and further to the level of 
	moduli 3-stacks, there combining it with the flux quantization condition.
	This involves a generalization of string orientifolds to what we 
	call {\it membrane orientifolds}. 
\end{enumerate}

In section \ref{Overview} we give an informal discussion of
central ideas of our constructions. 
In section \ref{Sec ing}
we recollect and set up the mathematical machinery needed. 
Then in section \ref{supergravityCField} 
discuss our model and analyze its properties.

%%%%%%%%%%%%%%%%%%%%%%%%%%%%%%%%%
\section{Informal overview}
\label{Overview}
%%%%%%%%%%%%%%%%%%%%%%%%%%%%%%%%

The following sections are written in formal mathematical style. 
But in order to provide the pure physicist reader with a working idea of what 
the formalization is about, and in order to help the 
pure mathematician reader get a working idea of the physical meaning
of the homotopy-theoretic constructions, we give in 
this section an informal discussion of some central ideas and of
our main construction (see also the Introduction of \cite{FiorenzaSatiSchreiberII}) .

\medskip

The ambient theory in which higher gauge theory is naturally
formulated is the combination of differential geometry
with homotopy theory: \emph{higher differential geometry}.\footnote{
See section 1.2 of \cite{survey} for a gentle introduction 
and section 3.3 for a detailed account.}
With hindsight, this has its very roots in gauge theory. A \emph{BRST complex}
with its ghost fields and ghosts-of-ghosts and so forth, up to ghosts or order $n$
is secretly a \emph{Lie $n$-algebroid}, the higher analog of a Lie algebra.\footnote{
See section 1.3.5 of \cite{survey} for a gentle introduction, and section 3.4 for a
detailed account.}
Whereas a Lie algebra encodes an infinitesimal symmetry of a single object, 
 a BRST complex encodes several objects -- the gauge field configurations --
together with the infinitesimal symmetries -- the gauge transformations -- between them,
together with the symmetries of symmetries between those, and so on.
Just as a Lie algebra is the approximation to a finite smooth object,
a Lie group, so a Lie $n$-algebroid is the approximation to a finite smooth
object: this is called a \emph{smooth $n$-groupoid} or, equivalently, 
a \emph{smooth $n$-stack}.
For instance, for $G$ a Lie group and $X$ a smooth manifold, there is a smooth stack
of $G$-gauge fields on $X$, 
which we denote $[X, \mathbf{B}G_{\mathrm{conn}}]$,
and which is the finite version of the BRST-complex of 
(off-shell) $G$-Yang-Mills theory on $X$.
If we forget the smooth structure on this, we write 
$\mathbf{H}(X, \mathbf{B}G_{\mathrm{conn}})$ 
for the underlying \emph{groupoid} of field configurations: it contains, 
as its \emph{objects}, the gauge field configurations, and, as its 
\emph{morphisms}, all the gauge transformations between these.
By quotienting out the gauge transformations we obtain the plain set 
$$
  \hat H^1(X,G)
  :=
  H(X, \mathbf{B}G_{\mathrm{conn}})  
$$
of gauge equivalence classes, which physically is the set of gauge equivalence classes
of $G$-gauge field configurations on $X$, and which mathematically is the
degree-1 \emph{nonabelian differential cohomology} on $X$ with coefficients in $G$.

\vspace{3mm}
The simplest example of interest is obtained for $G = U(1)$, in which case
$\mathbf{H}(X, \mathbf{B}U(1)_{\mathrm{conn}})$ is the groupoid
of Maxwell field configurations on $X$.
The examples of interest to us are $G = E_8$, the
largest exceptional simple Lie group, and $G={\rm Spin}$. 
In the first case,  
$\mathbf{H}(X, (\mathbf{B}E_8)_{\mathrm{conn}})$ is the groupoid of $E_8$-gauge
fields as they live, for instance, on a Ho{\v r}ava-Witten boundary of 11-dimensional
spacetime.
In the second case $G = \mathrm{Spin}$,  
$\mathbf{H}(X, \mathbf{B}\mathrm{Spin}_{\mathrm{conn}})$ is the groupoid
of $\mathrm{Spin}$-connections on $X$, which, in the first-order formulation of
gravity that is of relevance in supergravity, encodes part of the field of
gravity itself.

\vspace{3mm}
All these examples admit \emph{higher analogs}. For instance, for every
natural number $n$, there is a moduli $n$-stack of \emph{$n$-form gauge fields},
which we write $\mathbf{B}^n U(1)_{\mathrm{conn}}$.
This is such that $[X, \mathbf{B}^n U(1)_{\mathrm{conn}}]$ is the Lie integration
of the BRST complex of (off-shell) $n$-form field configurations.
Then $\mathbf{H}(X, \mathbf{B}^n U(1)_{\mathrm{conn}})$ is the underlying 
\emph{$n$-groupoid} of field configurations. Its objects are,
locally on patches $U \hookrightarrow X$, given by differential $n$-forms
$C_U$. Its gauge transformations between fields $C_U$ and $C'_U$ 
are locally given by $(n-1)$-forms $B_U$, such that 
$$
  C'_U = C_U + d B_U
  \,.
$$
Its gauge-of-gauge transformations between gauge transformations \{$B_U$, $B'_U$\}
are $(n-2)$-forms $A_U$, such that
$$
  B'_U = B_U + d A_U
  \,.
$$
The pattern continues in a similar fashion. 
The global structure is more intricate, but is essentially given by gluing such 
local data on intersections of patches by precisely such higher gauge transformations.

\vspace{3mm}
It is clear from the above discussion that the supergravity C-field is bound to be essentially
an object in $\mathbf{H}(X, \mathbf{B}^3 U(1)_{\mathrm{conn}})$. But the situation is
slightly more involved, because there is a quantum constraint on the C-field.
All we have to do is add this constraint to the picture, making sure this is done in the
proper gauge theoretic way.
More precisely,  the C-field interacts with the field of gravity,
whose configurations are $\mathbf{H}(X, \mathbf{B}\mathrm{Spin}_{\mathrm{conn}})$,
and, over Ho{\v r}ava-Witten boundaries, with an $E_8$-gauge field in 
$\mathbf{H}(\partial X, (\mathbf{B}E_8)_{\mathrm{conn}})$; this extends
to the
bulk, at least at the level of the underlying principal bundles  in 
$\mathbf{H}(X, \mathbf{B}E_8)$. Moreover, every Spin connection
and every $E_8$-connection
induces associated \emph{Chern-Simons circle 3-bundles} via maps\footnote{
  Details are in \cite{FSS}. See also section 4.1 of \cite{survey}.
} 
of 3-stacks denoted
$$
  \tfrac{1}{2}\hat {\mathbf{p}}_1 : \mathbf{B}\mathrm{Spin}_{\mathrm{conn}}
  \to
  \mathbf{B}^3 U(1)_{\mathrm{conn}}
$$
and
$$
  \hat {\mathbf{a}} : (\mathbf{B}E_8)_{\mathrm{conn}}
  \to
  \mathbf{B}^3 U(1)_{\mathrm{conn}}
  \,.
$$
The quantum constraint on these three fields (reviewed below in section \ref{CSWithBackgroundCharge})
 is on -- integral cohomology 
classes (``instanton sectors'') -- given by the equation (see \cite{WittenFluxQuantization})
\(
  [2G_4] = \tfrac{1}{2}p_1 + 2a
  \,,
  \label{QuantizationConditionInOverview}
\)
where $[G_4]$ is the class of the C-field.
It is useful to encode that equation graphically: the set of 
triples of gauge equivalence classes of fields that satisfies this equation is the
\emph{fiber product} or \emph{pullback} of the maps on cohomology sets
$$
  \tfrac{1}{2}\hat {{p}}_1  + 2 {a}
  : 
  H(X,\mathbf{B}\mathrm{Spin}_{\mathrm{conn}}
  \times
   \mathbf{B}E_8)_{\mathrm{conn}}
   )
  \to
  H(X,\mathbf{B}^3 U(1)_{\mathrm{conn}})
$$
and the map
$$
  2G_4
  :
  H(X, \mathbf{B}^3 U(1)_{\mathrm{conn}})
  \to 
  H(X, \mathbf{B}^3 U(1))
$$
that simply forgets the underlying connection data. 
Namely, the solution set of 
(\ref{QuantizationConditionInOverview}) on cohomology 
is the set that universally completes, in the top left corner, 
this square of functions between sets:
$$
  \raisebox{20pt}{
  \xymatrix{
     \ar[rr] 
	 \ar[d]
	  && H(X, \mathbf{B}^3 U(1)_{\mathrm{conn}})
	 \ar[d]^{2G_4}
	 \\
	 H(X, \mathbf{B}\mathrm{Spin}_{\mathrm{conn}} \times \mathbf{B}E_8)
	 \ar[rr]^{~~~\tfrac{1}{2} p_1 + 2 a}
	 && 
	 H(X, \mathbf{B}^3 U(1))~~.
  }
  }
$$
From this perspective it might seem as if imposing the
quantization condition simply restricts the set of possible 
field configurations to the subset of those triples that satisfy the 
quantization condition.
But a moment of reflection shows that this is wrong: physically, 
because for \emph{quantization}
we must not be working with sets of gauge equivalence classes of field
configurations. Instead, we need to retain at least 
the full BRST complexes of fields,
and better yet, as we do here, retain also the finite gauge transformations,
hence consider the $n$-groupoids of field configurations. 
Mathematically, the reason is that forming an ordinary fiber product
in homotopy theory breaks the universal property of the pullback 
and hence makes it useless, in fact meaningless.

\vspace{3mm}
We find that, in either case, implementing the above quantum 
constraint equation means forming a universal square as above,
but using the higher groupoids $\mathbf{H}(X,-)$ of field configurations, 
gauge transformations, and higher gauge transformation, instead
of just the gauge equivalence classes $H(X,-)$.
Doing so gives what mathematically is called 
forming a \emph{homotopy pullback} square: a square diagram
$$
  \xymatrix{
     \ar[rr]_>{\ }="s"
	 \ar[d]
	  && \mathbf{H}(X, \mathbf{B}^3 U(1)_{\mathrm{conn}} )
	 \ar[d]^{2\mathbf{G}_4}
	 \\
	 \mathbf{H}(X, \mathbf{B}\mathrm{Spin}_{\mathrm{conn}}\times \mathbf{B}E_8)
	 \ar[rr]_{~~~~\tfrac{1}{2}\mathbf{p}_1 + 2 \mathbf{a}}^<{\ }="t"
	 &&
	 \mathbf{H}(X, \mathbf{B}^3 U(1)) 
	 \ar@{=>}_\simeq "s"; "t"
  }
$$
of maps of higher groupoids, where now 
everything holds only \emph{up to gauge transformations}
or \emph{up to homotopy}, 
as indicated by the double arrow now filling this diagram. This is
the most natural thing to do physically: if condition
(\ref{QuantizationConditionInOverview}) is to hold for 
gauge equivalence classes of fields then, clearly, 
on the actual fields there is a 
gauge transformation exhibiting the equivalence. 

\vspace{3mm}
The mathematics of homotopy theory provides a calculus for handling
such constructions up to gauge transformations. Homotopy theory
is precisely the formalism for dealing with gauge systems and
higher gauge systems, and this is what we use in the following.
Accordingly, \emph{all} square diagrams as above appearing later
in this paper are implicitly filled by a gauge transformation, even
if we will usually suppress this from the notation.
Moreover, in this construction the choice of $X$ is not essential.
We may in full generality ask for the \emph{universal smooth moduli $n$-stack}
of C-field configurations, to be denoted $\mathbf{CField}$. This 
is to be such that for any manifold $X$, morphisms of smooth higher stacks
$$
   X \to \mathbf{CField}
$$
correspond precisely to triples of fields (gravity, C-field, $E_8$-field) 
on $X$, satisfying the quantization condition (\ref{QuantizationConditionInOverview}) 
up to a specified gauge equivalence, and such that 
homotopies between such maps correspond precisely to compatible gauge transformations
between such triples of field configurations.
By a basic but fundamental fact of higher geometry, this universal
moduli 3-stack is necessarily characterized as completing the analogous 
diagram as above, now consisting of fully fledged morphisms of higher smooth
stacks. In other words, the moduli 3-stack $\mathbf{CField}$ is 
to be this homotopy pullback of higher moduli stacks:
\(
  \raisebox{20pt}{
  \xymatrix{
    \mathbf{CField}
    \ar[rr]_>{\ }="s"
	\ar[d]
	&&
	\mathbf{B}^3 U(1)_{\mathrm{conn}}
	\ar[d]^{2\mathbf{G}_4}
	\\
    \mathbf{B}\mathrm{Spin}_{\mathrm{conn}}
	\times
	\mathbf{B}E_8
	\ar[rr]_{~~~\tfrac{1}{2}{\mathbf{p}}_1 + 2 \mathbf{a}}^<{\ }="t"
	&&
	\mathbf{B}^3 U(1)~~.
	\ar@{=>}_\simeq "s"; "t"
  }
  }
\)
In summary, this is the straightforward translation of the constraint equation
(\ref{QuantizationConditionInOverview}) from gauge equivalence classes
to genuine higher gauge field configurations. And this is the model
for the C-field that we present here. We show in the following sections that 
this construction
reproduces all the relevant properties of previous proposals and refines
them from 1-groupoids of fields and gauge-of-gauge equivalence classes
of gauge transformations to the full 3-groupoid of field configurations
and further to the full smooth moduli 3-stack of field configurations. 

\vspace{3mm}
The boundary data of C-field configurations in section \ref{CFieldRestrictionToTheBoundary}
is constructed analogously: the two physical conditions (that the $E_8$ gauge field
becomes dynamical and that the $C$-field class trivializes) have straightforward
translation into homotopy pullback diagrams. 
We show in the final prop. \ref{BoundaryConfigsAreString2aConnections} 
that the moduli 2-stack of C-field boundary conditions obtained this way 
is precisely that of anomaly-free heterotic field configurations as 
found in \cite{SSSIII}.

%%%%%%%%%%%%%%%%%%%%%%%%%%%%
\section{Ingredients}
%%%%%%%%%%%%%%%%%%%%%%%%%%%%
\label{Sec ing}

Before we come to our main constructions in section
\ref{supergravityCField} 
we briefly lay some foundations. 
First we recall in section \ref{HigherNobabelianGaugeTheory}
basics of smooth moduli stacks, of the refinement of 
ordinary \emph{abelian} differential cohomology to moduli stacks,
and then of those aspects of \emph{nonabelian} differential cohomology 
that we need in the following sections.
Then we recall in section \ref{CSWithBackgroundCharge} the 
origin of the factor of 2, that governs the whole discussion here,
from quadratic refinement of higher abelian Chern-Simons functionals.
Finally, in section \ref{DifferentialIntegralWuStructures},
we first review the formalization in \cite{HopkinsSinger}
of this situation in terms of \emph{differential integral Wu classes}
and then show how this refines to nonabelian differential cohomology.
This leads over seamlessly to the model of the C-field 
introduced further below in section \ref{supergravityCField}.

%%%%%%%%%%%%%%%%%%%%%%%%%%%%%%%%%%%%%%%%%%%%%%%%%%%%
\subsection{Abelian and nonabelian differential cohomology}
\label{HigherNobabelianGaugeTheory}
%%%%%%%%%%%%%%%%%%%%%%%%%%%%%%%%%%%%%%%%%%%%%%%%%%%%

We give a list of the basic definitions and properties of
\begin{enumerate}
  \item smooth higher groupoids / smooth higher stacks,
  \item abelian differential cohomology refined to smooth moduli stacks,
  \item nonabelian differential cohomology,
\end{enumerate}
that we invoke below in section \ref{supergravityCField}. This list
is necessarily somewhat terse. For a comprehensive account
we refer the reader to \cite{survey}. Much of the necessary technology is
spelled out in \cite{FSS}, and much of the relation to phenomena
in string theory is discussed in \cite{SSSIII} and in 
the companion article \cite{FiorenzaSatiSchreiberII}.

\medskip

Differential geometry can be viewed as the geometry modeled on the following \emph{site}.
\begin{definition}
  Let $\mathrm{CartSp}$ be the category whose objects are the Cartesian
  spaces $\mathbb{R}^n$ for $n \in \mathbb{N}$, and whose morphisms are
  the smooth functions between these.
  A family of morphism $\{U_i \to U\}$ in $\mathrm{CartSp}$ is
  called a \emph{good cover} if each all non-empty finite intersections
  of the $U_i$ in $U$ is diffeomorphic to a Cartesian space. 
  This defines a \emph{coverage} (pretopology)
  and we regard $\mathrm{CartSp}$ as a site equipped with this coverage.
\end{definition}
Higher differential geometry takes place in the \emph{$\infty$-topos}
over this site.\footnote{See section 2.1.4 and 3.3 in \cite{survey}.}
\begin{definition}
  Write 
  $$
    \mathrm{Smooth}\infty \mathrm{Grpd} := \mathrm{Sh}_\infty(\mathrm{CartSp})
  $$
  for the $\infty$-category of higher stacks over $\mathrm{CartSp}$. 
  As a simplicial category, this is the \emph{simplicial localization} $L_W$
  of the category of simplicial presheaves 
  $[\mathrm{CartSp}^{\mathrm{op}}, \mathrm{sSet}]$ 
  over $\mathrm{CartSp}$,  at the
  set $W$ of morphisms which are stalkwise weak homotopy equivalences of simplicial sets:
  $$
    \mathrm{Smooth}\infty \mathrm{Grpd}
	\simeq
  	L_W [\mathrm{CartSp}^{\mathrm{op}}, \mathrm{sSet}]
  $$
\end{definition}
\begin{remark}
  The localization formally inverts the morphisms in $W$ and is analogous to 
  the possibly more familiar localization at \emph{quasi-isomorphisms}
  that yields the derived categories of topological branes for the topological
  string.  Here we are dealing with a non-abelian generalization and
  refinement of this process. Instead of just quasi-isomorphisms between
  chain complexes we have more generally weak homotopy equivalences between simplicial sets,
  and the formal inverses that we add are just \emph{homotopy inverses}, but
  we also add the relevant homotopies, the relevant homotopies between homotopies, and so on.
\end{remark}
Usually we write 
$$
  \mathbf{H} := \mathrm{Smooth}\infty \mathrm{Grpd}
$$
for short, which is suggestive in view of the following
\begin{definition} 
  For $X,A \in \mathbf{H}$ any two higher stacks, the 
hom-$\infty$-groupoid between them is denoted $\mathbf{H}(X, A)$.
We also call this the \emph{cocycle $\infty$-groupoid} for
cocycles on $X$ with coefficients in $A$. 
For its set of connected components
we write
$$
  H(X,A) := \pi_0\mathbf{H}(X, A)
$$
and speak of the \emph{smooth nonabelian cohomology}
or just \emph{cohomology} set, for short, on $X$ with coefficients in $A$. 
\end{definition}
\begin{example}
  The following differential geometric objects are naturally embedded into
  $\mathbf{H}$:
  \begin{enumerate}
    \item smooth manifolds;
	\item smooth orbifolds;
	\item more general Lie groupoids / differentiable stacks;
	\item diffeological spaces, such as smooth mapping spaces $C^\infty(\Sigma,X)$
	between manifolds (e.g. sigma models);
	\item 
	smooth moduli stacks $\mathbf{B}G$ of $G$-principal bundles, for $G$ a Lie
	group;
	\item smooth moduli stacks $\mathbf{B}G_{\mathrm{conn}}$ and
	$\mathrm{Loc}(G) \simeq \mathbf{B}G_{\mathrm{flat}}$ of $G$-principal 
	bundles with connection and with flat connection, respectively.
  \end{enumerate}
  For the last two items see also Example \ref{DeloopingExamples} below.
  
  There are many more and ``higher'' examples. Some of these we describe in detail
  in the following.
\end{example}
We will need only some basic facts of $\infty$-category theory.\footnote{
Such as summarized in sections 2.1.1 and 2.1.2 of \cite{survey}, see the references
provided there.} 
One fundamental fact is the existence of all \emph{$\infty$-pullbacks} /
\emph{homotopy pullbacks} in $\mathbf{H}$. In section 2.1.4.2 of \cite{survey}
is a discussion of explicit constructions of these, which many of our computations
in the following rely on.
Another fundamental fact that we will use frequently is
\begin{proposition} [pasting law for homotopy pullbacks] 
  \label{PastingLawForPullbacks}
  Let 
  $$
    \xymatrix{
       A \ar[r] \ar[d] & B \ar[r]\ar[d] & C \ar[d]
       \\
       D \ar[r] & E \ar[r] & F
    }
  $$
  be a diagram in $\mathbf{H}$ and suppose that the right square is a 
  homotopy pullback. Then the left square is a homotopy pullback precisely if the 
  outer rectangle is.
\end{proposition}
Using this alone there is induced a notion of higher group objects in $\mathbf{H}$.
\begin{definition}
  Write $\infty \mathrm{Grp}(\mathbf{H})$ for the $\infty$-category
  of group $\infty$-stacks  over
  $\mathrm{CartSp}$ (grouplike smooth $A_\infty$-spaces). 
  We call these \emph{smooth $\infty$-groups}.\footnote{
  See section 2.3.2 of \cite{survey}.}
  Write $\mathbf{H}^{*/}$ for the $\infty$-category of pointed objects in $\mathbf{H}$.
  Write 
  $$
    \Omega : \mathbf{H}^{*/} \to \infty \mathrm{Grp}(\mathbf{H})
  $$
  for the $\infty$-functor that sends a pointed object $* \to A$ to 
  its loop space object $\Omega_* A$, defined to be the homotopy pullback
  $$
    \raisebox{20pt}{
    \xymatrix{
	   \Omega_* A \ar[r] \ar[d] & {*} \ar[d]
	   \\
	   {*} \ar[r] & A
	}
	}
	\,.
  $$
\end{definition}
\begin{proposition}
  $\mathbf{H}$ has \emph{homotopy dimension} 0, hence every
  connected object $A$ has a point $* \to A$
  (necessarily essentially unique).
\end{proposition}
\begin{theorem} [Lurie]
  Looping induces an equivalence of $\infty$-categories
  $$
    \xymatrix{
      \infty \mathrm{Grp}(\mathbf{H})
	    \ar@{<-}@<+5pt>[rr]^{\hspace{2mm}\Omega}
	    \ar@<-5pt>[rr]^{~~\simeq}_{~~\mathbf{B}}
	  &&
	  \mathbf{H}^{{*}/}_{\geq 1}
	}
  $$
  between smooth $\infty$-groups and pointed \emph{connected} objects.
  The homotopy inverse functor $\mathbf{B}$ we call the \emph{delooping}
  functor.
\end{theorem}
\begin{definition}
If an $\infty$-stack $A$ is an $n$-fold loop object, we write $\mathbf{B}^n A$
for its $n$-fold delooping. For $X$ any other object we write
$$
  H^n(X,A) := \pi_0 \mathbf{H}(X, \mathbf{B}^n A)
$$
and speak of the \emph{degree $n$ cohomology} on $X$ with coefficients in $A$.
\end{definition}
\begin{example}
  Every Lie group $G$ is naturally also a smooth $\infty$-group. 
  Its delooping $\mathbf{B}G$ is the \emph{moduli stack} of $G$-principal
  bundles: for any smooth manifold $X$, the cocycle groupoid 
  $$
    \mathbf{H}(X, \mathbf{B}G)
	\simeq
	G \mathrm{Bund}(X)
  $$
  is the groupoid of smooth $G$-principal bundles 
  and smooth gauge transformations between them, on $X$. The corresponding nonabelian
  smooth cohomology 
  $$
    H^1(X, G) := H(X, \mathbf{B}G)
  $$
  coincides with degree-1 nonabelian {\v C}ech cohomology on $X$
  with coefficients in the sheaf of smooth $G$-valued functions.
  \label{DeloopingExamples}
  
  \vspace{3mm}
   If $G$ is an \emph{abelian} Lie group, such as $G = U(1)$,
  the delooping moduli stack $\mathbf{B}U(1)$ is itself again canonically an $\infty$-group,
  called the \emph{circle 2-group}. In fact for all $n \in \mathbb{N}$
  the $n$-fold delooping $\mathbf{B}^n U(1)$ exists. This is 
  the \emph{moduli $n$-stack for circle $n$-bundles}. 
  Morphisms $X \to \mathbf{B}^2 U(1)$ may be identified with \emph{bundle gerbes}
  on $X$ (circle 2-bundles), morphism $X \to \mathbf{B}^3 U(1)$ with \emph{bundle 2-gerbes}
  (circle 3-bundles) and so on.
  The smooth cohomology
  $$
    H^n(X, U(1)) := H(X, \mathbf{B}^n U(1))
  $$
  coincides with degree-$n$ {\v C}ech cohomology with coefficients in 
  the sheaf of smooth $U(1)$-valued functions.
  
  \vspace{3mm}
  Generally, for $A$ a sheaf of abelian groups, 
  $$
    H^n(X, A) := H(X, \mathbf{B}^n A) := \pi_0 \mathbf{H}(X, \mathbf{B}^n A)
  $$
  coincides with the \emph{sheaf cohomology} in degree $n$ over $X$ with 
  coefficients in $A$.
\end{example}
But see also the further example \ref{RelationSingularSmoothCohomology} below.
\begin{proposition}
  \label{GeometricRealization}
  The inclusion
  $$
    \mathrm{Disc} : 
	 \xymatrix{
	    \mathrm{Top} 
		\ar[r]^{\hspace{-4mm}\mathrm{Sing}}_{\hspace{-4mm}\simeq}
		&
		\infty \mathrm{Grpd}~
		\ar@{^{(}->}[r]
        &
		\mathbf{H}
	}
  $$
  of topological spaces into smooth higher stacks -- as the 
  \emph{discrete} or \emph{locally constant} smooth $\infty$-stacks --
  has a derived left adjoint
  $$
    \vert -\vert 
	   : 
	\xymatrix{
	   \mathbf{H}
       \ar[r]^{\hspace{-5mm}\Pi}
       &		
	   \infty\mathrm{Grpd} 
	   \ar[r]^{~\vert - \vert}_{~\simeq}
	   &
	   \mathrm{Top}
	}
	\,,
  $$
  called the \emph{geometric realization} of smooth higher stacks.
\end{proposition}
%See in \cite{survey} section 3.2.2 and 3.3.3.
\begin{proposition}
  If a higher group $G$ has a presentation as a simplicial 
  presheaf which in turn is presented by a well-pointed simplicial 
  topological group that is degreewise paracompact, then 
  for $X$ any manifold, we have an isomorphism
  $$
    H^1(X, G)
	\simeq
    \pi_0\mathbf{H}(X, \mathbf{B}G)
	\simeq
	\pi_0 \mathrm{Top}(X, \vert B  G\vert)
	\simeq
	\pi_0 \mathrm{Top}(X,  B  \vert G\vert)
  $$
  of the smooth higher nonabelian cohomology of $X$ with coefficients
  in $G$ and homotopy classes of maps into the geometric realization of
  the higher moduli stack.
\end{proposition}
This follows by \cite{RobertsStevenson}. See section 3.2.2 of \cite{survey}.
\begin{remark}
  In terms of gauge theory this says that 
  for $G$ a higher group, the geometric realization
  $\vert \mathbf{B}G\vert$ is the classification space of the
  instanton sector of higher $G$-gauge field configurations.
\end{remark}
\begin{definition}
  By general facts, $\mathrm{Disc} : \infty \mathrm{Grpd} \hookrightarrow \mathbf{H}$
  is also itself a derived left adjoint. For $A \in \mathbf{H}$ any object,
  we write $\flat A \to A$ for the counit of the corresponding adjunction.
  
  \vspace{3mm}
 For $G$ an $\infty$-group, we call $\flat \mathbf{B}G$ the higher moduli stack of
  \emph{flat $G$-princial $\infty$-connections} or of 
  \emph{$G$-local systems}.\footnote{See 2.3.12 in \cite{survey}.}
\end{definition}
\begin{example}
  The moduli $n$-stack $\flat \mathbf{B}^n U(1)$ is presented by
  the complex of sheaves concentrated in degree 1 on the 
  \emph{constant} sheaf with values $U(1)$. This may be thought of
  as the sheaf of functions into the \emph{discrete} group 
  $U(1)_{\mathrm{disc}}$ underlying the Lie group
  $U(1)$:
  $$
    \mathbf{B}^n U(1)_{\mathrm{disc}}  \simeq \flat \mathbf{B}^n U(1) 
	\,.
  $$
  The smooth cohomology with coefficients in this discrete object 
  coincides with ordinary \emph{singular cohomology} with coefficients in $U(1)$
  $$
    H^n(X, U(1)_{\mathrm{disc}}) \simeq
	H(X, \flat \mathbf{B}^n U(1))
	\simeq
	H^n_{\mathrm{sing}}(X, U(1))
	\,.
  $$
  \label{RelationSingularSmoothCohomology}
\end{example}
%This is discussed in 3.3.9 of \cite{survey}.
\begin{definition}
  For $G$ a smooth $\infty$-group, 
  we write
  $$
    \mathbf{\flat}_{\mathrm{dR}} \mathbf{B}G 
	  :=
	\flat \mathbf{B}G \times_{\mathbf{B}G} {*}
  $$
  for the homotopy pullback of the counit $\flat \mathbf{B}G \to \mathbf{B}G$ 
  along the point inclusion. 
  Smooth cohomology with coefficients in $\flat_{\mathrm{dR}}\mathbf{B}G$
  we call \emph{$G$-de Rham cohomology}.
  The canonical morphism
  $$
    \theta_G : G \to \mathbf{\flat}_{\mathrm{dR}}\mathbf{B}G
  $$
  we call the \emph{Maurer-Cartan form} on the smooth $\infty$-group $G$.
  Specifically for $G = \mathbf{B}^n U(1)$ the circle $(n+1)$-group,
  we also write 
  $$
    \mathrm{curv}
	:=
	\theta_{\mathbf{B}^n U(1)} : \mathbf{B}^n U(1) 
	\to \mathbf{\flat}_{\mathrm{dR}}\mathbf{B}^{n+1}U(1)
  $$
  and speak of the \emph{universal curvature characteristic map} in degree $(n+1)$.
  \label{UniversalCurvatureCharcteristic}
\end{definition}
%See 2.3.14 and 2.3.16 of \cite{survey}.
\begin{proposition}
  Under the Dold-Kan correspondence \footnote{See 2.1.7 of \cite{survey}}, 
  $\flat_{\mathrm{dR}} \mathbf{B}^n U(1)$
  is presented by the truncated de Rham complex of sheaves of abelian groups
  $$
    \xymatrix{
      \Omega^1(-)
      \ar[r]^{d_{\mathrm{dR}}}	  
	  &
	  \Omega^2(-)
      \ar[r]^{d_{\mathrm{dR}}}	  
	  &
	  \cdots
      \ar[r]^{\hspace{-3mm}d_{\mathrm{dR}}}	  
	  &
	  \Omega^n_{\mathrm{cl}}(-)
	}
	\,.
  $$
  Morover, the universal curvature characteristic $\mathrm{curv}$
  is presented by a correspondence of simplicial presheaves
  $$
    \raisebox{20pt}{
    \xymatrix{
	  \mathbf{B}^n U(1)_{\mathrm{diff}}
	  \ar@{->>}[d]^\wr
	  \ar[r]^{\hspace{-3mm}\mathrm{curv}}
	  &
	  \flat_{\mathrm{dR}}\mathbf{B}^{n+1} U(1)
      \\
	  \mathbf{B}^n U(1)
	}
	}
	\,,
  $$
  where $\mathbf{B}^n U(1)_{\mathrm{conn}}$
  classifies circle $n$-bundles equipped with \emph{pseudo-connection}:
  they carry connection data, but gauge transformations are allowed
  to freely shift the connections.
  \label{PresentationOfUniversalCurvature}
\end{proposition}
%See 3.3.10 of \cite{survey}.
\begin{definition}
  For $n \geq 1$, the \emph{moduli $n$-stack of circle $n$-bundles with connection}
  $\mathbf{B}^n U(1)_{\mathrm{conn}}$
  is the homotopy pullback of higher stacks
  $$
  \xymatrix{
    \mathbf{B}^n U(1)_{\mathrm{conn}}
    \ar[d]
    \ar[r]^{}
    & \Omega^{n+1}_{\mathrm{cl}}
    \ar[d]
    \\
    \mathbf{B}^n U(1)
    \ar[r]^{\hspace{-3mm}\mathrm{curv}}
    &
    \mathbf{\flat}_{\mathrm{dR}} \mathbf{B}^{n+1}\mathbb{R}~~. 
  }
  $$
  \label{BnU1conn}  
\end{definition}
%See 2.3.17 of \cite{survey}.
\begin{proposition}
  Under the Dold-Kan correspondence $\mathbf{B}^n U(1)_{\mathrm{conn}}$
  is presented by the Beilinson-Deligne complex of sheaves,
  either in the form  
  $$
    \xymatrix{
	  C^\infty(-, U(1))~~
	  \ar[r]^{~~~d_{\mathrm{dR}}\mathrm{log}}
	  &
      \Omega^1(-)
      \ar[r]^{d_{\mathrm{dR}}}	  
	  &
	  \Omega^2(-)
      \ar[r]^{d_{\mathrm{dR}}}	  
	  &
	  \cdots
      \ar[r]^{\hspace{-2mm}d_{\mathrm{dR}}}	  
	  &
	  \Omega^n(-)
	}
  $$
  or equivalently in the form
  $$
    \xymatrix{
	  \mathbb{Z}
	  \ar[r]
	  &
	  C^\infty(-, \mathbb{R})
	  \ar[r]^{~~d_{\mathrm{dR}}}
	  &
      \Omega^1(-)
      \ar[r]^{d_{\mathrm{dR}}}	  
	  &
	  \Omega^2(-)
      \ar[r]^{d_{\mathrm{dR}}}	  
	  &
	  \cdots
      \ar[r]^{\hspace{-2mm}d_{\mathrm{dR}}}	  
	  &
	  \Omega^n(-)
	}
	\,.
  $$
  For $X$ a smooth manifold, the corresponding cohomology
  $$
    H(X, \mathbf{B}^n U(1)_{\mathrm{conn}})
	\simeq
	\hat H^{n+1}(X)
  $$  
  is the \emph{ordinary differential cohomology} of $X$ in degree
  $(n+1)$.
  \label{DoldKanOnDeligne}
\end{proposition}
%See 3.3.13 in \cite{survey}.
\begin{definition}
  For $G$ a topological group and
  $[c] \in H^{n+1}(B G, \mathbb{Z})$ a universal characteristic class,
  we say that a \emph{smooth refinement} of $[c]$ is a
  morphism of higher smooth moduli stacks of the form
  $$
    \mathbf{c} : \mathbf{B}G \to \mathbf{B}^n U(1)
  $$
  such that under geometric realization this 
  reproduces $c$, in that 
  $$
    \vert \mathbf{c} \vert : B G \to K(\mathbb{Z}, n+1)
  $$
  is a representative of $[c]$. 
  We say a further \emph{differential refinement} of $[c]$
  is a morphism of higher moduli stacks of the form
  $$
    \hat {\mathbf{c}} : 
	\mathbf{B}G_{\mathrm{conn}}
	\to
	\mathbf{B}^n U(1)_{\mathrm{conn}}
  $$
  such that it completes the diagram
  $$
    \raisebox{20pt}{
    \xymatrix{
	  \mathbf{\flat} \mathbf{B}G 
	  \ar[r]^{\hspace{-3mm}\mathbf{\flat} \mathbf{c}}
	  \ar[d]
	  &
	  \mathbf{\flat}\mathbf{B}^n U(1)
	  \ar[d]
	  \\
   	  \mathbf{B}G_{\mathrm{conn}}
	  \ar@{-->}[r]^{\hspace{-3mm}\hat {\mathbf{c}}}
	  \ar[d]
	  &
	  \mathbf{B}^n U(1)_{\mathrm{conn}}
	  \ar[d]
	  \\
	  \mathbf{B}G
	  \ar[r]^{\hspace{-5mm}\mathbf{c}}
	  &
	  \mathbf{B}^n U(1)~~.
	}
	}
  $$
  \label{SmoothAndDifferentialRefinement}
\end{definition}
%See \cite{survey} 2.3.18. 
\begin{remark}
  A smooth refinement of $[c]$ is equivalently a smooth circle $n$-bundle
  / $(n-1)$-bundle gerbe on a smooth moduli stack whose higher Dixmier-Douady class
  is $[c]$. Similarly a differential refinement of $[c]$ is
  a circle $n$-bundle \emph{with connection} on the differential moduli stack
  whose Dixmier-Douady class is $[c]$.
\end{remark}
Using the $L_\infty$-algebraic data provided in 
\cite{SSSI}, the following was shown in \cite{FSS}. See also section 4.1 in \cite{survey}.
\begin{proposition}
  There exists a smooth and differential refinement of the first fractional 
  Pontrjagin class 
  $$
    \tfrac{1}{2}p_1 \in H^4(B \mathrm{Spin}, \mathbb{Z})
  $$ 
  to the smooth moduli stack of Spin connections
  with values in the smooth moduli 3-stack of circle 3-bundles with 
  3-connection
  $$
    \tfrac{1}{2} \hat {\mathbf{p}}_1 
	  :
    \xymatrix{	  
  	  \mathbf{B}\mathrm{Spin}_{\mathrm{conn}}
	  \ar[r]
	  &
	  \mathbf{B}^3 U(1)_{\mathrm{conn}}
	}
	\,.
  $$
  \label{FirstFracPontryagin}
  \label{FirstFractionalDifferentialPontrjagin}
\end{proposition}
%%%%%%%%%%%%%%%%%%%%%%%%%
%% the class on B E8   %%
%%%%%%%%%%%%%%%%%%%%%%%%%
\begin{proposition}
  Let $E_8$ be the largest semisimple exceptional Lie group.
  There exists a differential refinement of the canonical class
  $$
    [a] \in H^4(B E_8, \mathbb{Z})
  $$
  to the smooth moduli stack of $E_8$-connections
  with values in the smooth moduli 3-stack of circle 3-bundles with 
  3-connection
  $$
    \hat {\mathbf{a}}
	  :
    \xymatrix{	  
  	  (\mathbf{B}E_8)_{\mathrm{conn}}
	  \ar[r]
	  &
	  \mathbf{B}^3 U(1)_{\mathrm{conn}}
	}
	\,.
  $$
  \label{aDifferentially}
\end{proposition}
\begin{proposition}
  Under geometric realization, prop. \ref{GeometricRealization},
  the smooth class $\mathbf{a}$ becomes an equivalence
  $$
    \vert \mathbf{a}\vert
	: 
	B E_8
	\simeq_{16}
	B^3 U(1)
	\simeq 
	K(\mathbb{Z},4)
  $$
  on 16-coskeleta.
  \label{aIsMaxCompactSubgroup}
\end{proposition}
\proof
  %By \cite{BottSamelson} 
  The 15-coskeleton of the topological space
  $E_8$ is a $K(\mathbb{Z}, 4)$.
  By \cite{FSS}, $\mathbf{a}$ is a smooth refinement of
  the generator $[a] \in H^4(B E_8, \mathbb{Z})$. 
  By the Hurewicz theorem this is identified with 
  $\pi_4 (B E_8) \simeq \mathbb{Z}$. Hence in cohomology
  $\mathbf{a}$ induces an isomorphism
  $$
    \xymatrix@C=5pt{
	  \pi_4(B E_8)
      \simeq     
	  [S^4, B E_8]  
	  \simeq
	  H^1(S^4, E_8) 
      \ar[rrr]^{\vert \mathbf{a} \vert}|\simeq	  
	  &&&
	  H^4(S^4, \mathbb{Z}) 
	  \simeq
	  [S^4, K(\mathbb{Z},4)]
	  \simeq
	  \pi_4(S^4)
	  }
	\,.
  $$  
  Therefore $\vert \mathbf{a} \vert$ is a weak homotopy equivalence on 16 coskeleta.
\endofproof
%%%%%%%%%%%%%%%%%%%%%%%%%
%%   further stuff     %%
%%%%%%%%%%%%%%%%%%%%%%%%%
\begin{remark}
  We obtain the \emph{de Rham images} of these differential classes
  by postcomposition with the universal 4-curvature characteristic
  from def. \ref{UniversalCurvatureCharcteristic}:
  \bea
    (\tfrac{1}{2}\mathbf{p}_1)_{\mathrm{dR}}
	&:&
	\xymatrix{
	  \mathbf{B}\mathrm{Spin}_{\mathrm{conn}}
	  \ar[r]^{\hspace{-1mm}\hat {\mathbf{p}}_1}
	  &
	  \mathbf{B}^3 U(1)_{\mathrm{conn}}
	  \ar[r]^{\hspace{-2mm}\mathrm{curv}}
	  &
	  \mathbf{\flat}_{\mathrm{dR}} \mathbf{B}^4 U(1)~,
	}
  \nonumber\\
    \mathbf{a}_{\mathrm{dR}}
	&:&
	\xymatrix{
	  (\mathbf{B}E_8)_{\mathrm{conn}}
	  \ar[r]^{\hspace{-2mm}\hat {\mathbf{a}}}
	  &
	  \mathbf{B}^3 U(1)_{\mathrm{conn}}
	  \ar[r]^{\hspace{-2mm}\mathrm{curv}}
	  &
	  \mathbf{\flat}_{\mathrm{dR}} \mathbf{B}^4 U(1)~.
	}
	\nonumber
  \eea
  By prop. \ref{PresentationOfUniversalCurvature} these morphisms
  have a presentation by correspondences of simplicial presheaves
  $$
    \raisebox{20pt}{
    \xymatrix{
	  (\mathbf{B}E_8)_{\mathrm{diff}}
	  \ar@{->>}[d]^\wr
	  \ar[r]^{\hspace{-2mm}\mathbf{a}_{\mathrm{diff}}}
	  &
	  \mathbf{B}^3 U(1)_{\mathrm{diff}}
	  \ar[r]^{\mathrm{curv}}
	  \ar@{->>}[d]^\wr
	  &
	  \flat_{\mathrm{dR}}\mathbf{B}^4 \mathbb{R}
      \\
	  \mathbf{B}E_8
	  \ar[r]^{\hspace{-2mm}\mathbf{a}}
	  &
	  \mathbf{B}^3 U(1)
	}
	}
  $$
  involving the simplicial presheaf
  $(\mathbf{B}E_8)_{\mathrm{diff}}$ of $E_8$-\emph{pseudo-connections}.
  See \cite{FSS} for a thorough discussion.
  \label{DeRhamImagesOfDifferentialRefinements}
\end{remark}
Every morphism $\mathbf{c} : P \to B$ of higher pointed
stacks with homotopy fiber $A \to P$
may be regarded as
an \emph{$\infty$-bundle} over $B$ with typical fiber $A$. 
We may therefore consider the cohomology with coefficients in $A$
but \emph{twisted} by cocycles $\chi \in \mathbf{H}(X, B)$.\footnote{
See section 2.3.5 of \cite{survey}.}
Such an \emph{$\chi$-twisted} $A$-cocycle is a homotopy section
$\sigma$ in
$$
  \raisebox{20pt}{
  \xymatrix{
    & P
	\ar[d]^{\mathbf{c}}
    \\
    X 
	  \ar[r]
	  \ar@{-->}[ur]^\sigma
	& B \;.
  }
  }
$$
In the special case that $\mathbf{c}$ is interpeted as a smooth universal characteristic 
map as above, we think of a $\mathbf{c}$-twisted $A$-cocycle also as
a \emph{twisted $\mathbf{c}$-structure}.
\begin{definition}
  \label{TwistedCStructures}
  \index{twisted cohomology!twisted $\mathbf{c}$-structures}
  For $\mathbf{c} : \mathbf{B}G \to \mathbf{B}^n U(1)$ a smooth characteristic map
  in $\mathbf{H}$, define for any
  $X \in \mathbf{H}$ the
  $\infty$-groupoid $\mathbf{c}\mathrm{Struc}_{\mathrm{tw}}(X)$
  of \emph{twisted $\mathbf{c}$-structures}
  to be the $\infty$-pullback
  $$
    \xymatrix{
	  \mathbf{c}\mathrm{Struc}_{\mathrm{tw}}(X)
	  \ar[r]^{\mathrm{tw}} 
	  \ar[d]
	  & H^n(X, U(1))
	  \ar[d]
	  \\
	  \mathbf{H}(X, \mathbf{B}G)
	  \ar[r]^{\hspace{-5mm}\mathbf{c}}
	  &
	  \mathbf{H}(X, \mathbf{B}^n U(1))~,
	}
  $$
  where the vertical morphism on the right is the essentially unique
  effective epimorphism that picks a point in every connected component.

\vspace{3mm}
  For $\chi \in \mathbf{H}(X, \mathbf{B}^n A)$ a fixed twisting cocycle,
  the $\infty$-groupoid of \emph{$\chi$-twisted $\mathbf{c}$-structures}
  is the homotopy fiber
  $$
    \raisebox{20pt}{
    \xymatrix{
	  \mathbf{c}\mathrm{Struc}_{[\chi]}(X)
	  \ar[r]
	  \ar[d]
	  &
	  {*}
	  \ar[d]^{\chi}
	  \\
	  \mathbf{c}\mathrm{Struc}_{\mathrm{tw}}(X)
	  \ar[r]^{\hspace{-2mm}\mathrm{tw}}
	  &
	  H^n(X, U(1))~~.
	}
	}
  $$
\end{definition}
In \cite{SSSIII} (see also 4.4.4 in \cite{survey}) there is a list of examples of 
such nonabelian twisted cohomology governing anomaly 
cancellation in string theory: twisted $\mathrm{Spin}^c$ structures,
smooth twisted $\mathrm{String}$ structures, 
smooth twisted $\mathrm{Fivebrane}$ structures. 
The twisted String structures we re-encounter below 
in section \ref{CFieldRestrictionToTheBoundary} in the 
boundary field configurations of the C-field.

%%%%%%%%%%%%%%%%%%%%%%%%%%%%%%%%%%%%%%%%%%%%%%%%%%%%%%%%%%%%%%%%%%%%%%% 
\subsection{Higher abelian Chern-Simons theories with background charge} 
\label{CSWithBackgroundCharge}
%%%%%%%%%%%%%%%%%%%%%%%%%%%%%%%%%%%%%%%%%%%%%%%%%%%%%%%%%%%%%%%%%%%%%%% 

The supergravity C-field is an example of a general phenomenon of 
higher abelian Chern-Simons QFTs in the presence of 
\emph{background charge}. This phenomenon was originally noticed
in \cite{Witten96} and then made precise in \cite{HopkinsSinger}.
The holographic dual of this phenomenon is that of
self-dual higher gauge theories, which for the supergravity C-field
is the nonabelian 2-form theory on the M5-brane \cite{FiorenzaSatiSchreiberII}, 
and in this dual form it has been studied systematically in 
\cite{DFM,BM}.
We now review the idea in a way that will smoothly lead over to our
refinements to nonabelian higher gauge theory in  
section \ref{supergravityCField}.

\medskip

Fix some natural number $k \in \mathbb{N}$ and an oriented manifold
(compact with boundary) $X$ of dimension $4 k + 3$. 
The gauge equivalence class of a $(2k+1)$-form gauge field $\hat G$ on $X$ is an
element in the differential cohomology group $\hat H^{2k+2}(X)$.
The cup product $\hat G \cup \hat G \in \hat H^{4k+4}(X)$ of this class with itself 
has a natural higher holonomy over $X$, denoted
\bea
  \exp(i S (-)) : \hat H^{2k+2}(X) &\to & U(1)
\nonumber\\
  \hat G ~~ &\mapsto & \exp(i \int_X \hat G \cup \hat G)
  \,.
  \label{Eq action}
\eea
This is the exponentiated action functional for bare $(4k+3)$-dimensional
abelian Chern-Simons theory. For $k = 0$ this reduces to ordinary 3-dimensional
abelian Chern-Simons theory \cite{CS}. 
Notice that, even in this case, this is a bit more subtle
that Chern-Simons theory for a simply-connected gauge group $G$. In the latter
case all fields can be assumed to be globally defined forms. But in the non-simply-connected
case of $U(1)$, instead the fields are in general cocycles in differential
cohomology. If, however, we restrict attention to fields $C$ in the inclusion
$H^{2k+1}_{\mathrm{dR}}(X) \hookrightarrow \hat H^{2k+2}(X)$, then on these the above
action \eqref{Eq action} reduces to the familiar expression
$$
  \exp(i S(C)) = \exp(i \int_X C \wedge d_{\mathrm{dR} } C)
  \,.
$$
Observe now that
the above action functional may be regarded as a \emph{quadratic form}
on the group $\hat H^{2k+2}(X)$. The corresponding bilinear form is the 
(``secondary'', since $X$ is of dimension $4k+3$ instead of $4k+4$)
\emph{intersection pairing}
\bea
  \langle -,-\rangle : \hat H^{2k+2}(X) \times \hat H^{2k+2}(X) &\to & U(1)
\nonumber\\
  (\hat a_1~ , ~\hat a_2)\hspace{1cm} &\mapsto & \exp(i \int_X \hat a_1 \cup \hat a_2 )
  \,.
  \nonumber
\eea
However, note that from $\exp(i S(-))$ we do \emph{not} obtain a \emph{quadratic refinement} 
of the pairing. A quadratic refinement is, by definition, a function
$$
  q : \hat H^{2k+2}(X) \to U(1)
$$
(not necessarily homogenous of degree 2 as $\exp(i S(-))$ is), for which
the intersection pairing is obtained via the polarization formula
$$
  \langle \hat a_1, \hat a_2\rangle 
    = 
  q(\hat a_1 + \hat a_2)
  q(\hat a_1)^{-1}
  q(\hat a_2)^{-1}
  q(0)
  \,.
$$
If we took $q := \exp(i S(-))$, then the above formula would yield not 
$\langle -,-\rangle$, but the square $\langle -,-\rangle^2$, given by
the exponentiation of \emph{twice} the integral. 

\vspace{3mm}
The observation in \cite{Witten96} was 
that for the correct holographic physics, we need instead an action functional
which is indeed a genuine quadratic refinement of the intersection pairing.
But since the differential classes in $\hat H^{2k+2}(X)$ refine 
\emph{integral} cohomology, we cannot in general simply divide by 2 and
pass from $\exp( i \int_X \hat G \cup \hat G)$ to 
$\exp( i  \int_X \frac{1}{2} \hat G \cup \hat G)$. The integrand in the
latter expression does not make sense in general in differential cohomology.
If one tried to write it out in the ``obvious'' local formulas one would
find that it is a functional on fields which is not gauge invariant.
The analog of this fact is familiar from nonabelian $G$-Chern-Simons theory
with simply-connected $G$, where also the theory is consistent only at 
interger \emph{levels}. The ``level'' here is nothing but the 
underlying integral class $G \cup G$. 
Therefore, the only way to obtain a square root of the quadratic form
$\exp(i S(-))$ is to \emph{shift it}. Here we think of 
the analogy with a
quadratic form 
$$
  q : x \mapsto x^2
$$ 
on the real numbers (a parabola in the plane).
Replacing this by 
$$
  q^{\lambda} : x \mapsto x^2 - \lambda x
$$ 
for some real number
$\lambda$ means keeping the shape of the form, but shifting its minimum from 0 to
$\frac{1}{2}\lambda$. If we think of this as the potential term for a scalar field
$x$ then its ground state is now at $x = \frac{1}{2}\lambda$. We may say that there is
a \emph{background field} or \emph{background charge} that pushes
the field out of its free equilibrium.
See \cite{Freed,DFM,FM}.

\vspace{3mm}
To lift this reasoning to the action quadratic form 
$\exp(i S(-))$ on differential cocycles, we need a 
differential class $\hat \lambda \in H^{2k+2}(X)$ such that 
for every $\hat a \in H^{2k+2}(X)$ the composite class
$$
  \hat a \cup \hat a - \hat a \cup \hat \lambda
  \in 
  H^{4k+4}(X)
$$
is even, hence is divisible by 2. Because then we could define a shifted
action functional
$$
  \exp(i S^\lambda(-)) : \hat a \mapsto \exp\left(
  i \int_X \frac{1}{2}(\hat a \cup \hat a - \hat a \cup \hat \lambda)\right)
  \,,
$$
where now the fraction $\frac{1}{2}$ in the integrand does make sense. 
One directly sees that if this exists, then this shifted action is indeed a
quadratic refinement of the intersection pairing:
$$
  \exp(i S^\lambda(\hat a + \hat b))
  \exp(i S^\lambda(\hat a))^{-1}
  \exp(i S^\lambda(\hat b))^{-1}
  \exp(i S^\lambda(0))
  = 
  \exp(i \int_X \hat a \cup \hat b)
  \,.
$$
The condition on the existence of $\hat \lambda$ here means, equivalently, that the
image of the underlying integral class vanishes under the map
$$
  (-)_{\mathbb{Z}_2} : H^{2k+2}(X, \mathbb{Z}) \to H^{2k+2}(X, \mathbb{Z}_2)
$$
to $\mathbb{Z}_2$-cohomology:
$$
   (a)_{\mathbb{Z}_2} \cup (a)_{\mathbb{Z}_2} 
     - 
   (a)_{\mathbb{Z}_2} \cup (\lambda)_{\mathbb{Z}_2}  
   = 
   0 \in H^{4k+4}(X, \mathbb{Z}_2)
  \,.
$$
Precisely such a class $(\lambda)_{\mathbb{Z}_2}$ does uniquely exist on 
every oriented manifold. It is called the \emph{Wu class} 
$\nu_{2k+2} \in H^{2k+2}(X,\mathbb{Z}_2)$, and may be \emph{defined} by this condition.
Moreover, if $X$ is a $\mathrm{Spin}$-manifold, then every second
Wu class, $\nu_{4k}$, has a pre-image in integral cohomology, 
hence $\lambda$ does exist as required above
$$
  (\lambda)_{\mathbb{Z}_2}
    = 
  \nu_{2k+2}
  \,.
$$
It is given by 
polynomials in the Pontrjagin classes of $X$
(discussed in section E.1 of \cite{HopkinsSinger}). For instance
the degree-4 Wu class (for $k = 1$) is refined by the first fractional Pontrjagin class
$\frac{1}{2}p_1$
$$
  (\tfrac{1}{2}p_1)_{\mathbb{Z}_2} = \nu_4
  \,.
$$
In the present context, this was observed in \cite{Witten96} (see around eq. (3.3) there).

\vspace{3mm}
Notice that the equations of motion of the shifted action $\exp(i S^\lambda(\hat a))$
are no longer $\mathrm{curv}(\hat a) = 0$, but are now 
$$
  \mathrm{curv}(\hat a) =  \tfrac{1}{2}\mathrm{curv}(\hat \lambda)
  \,.
$$
We therefore think of
$\exp(i S^\lambda(-))$ as the exponentiated action functional for
\emph{higher dimensional abelian Chern-Simons theory with background charge 
$\frac{1}{2}\lambda$}.
With respect to the shifted action functional it makes sense to introduce the
shifted field
$$
  \hat G  := \hat a  - \tfrac{1}{2}\hat \lambda
  \,.
$$
This is simply a re-parameterization such that the 
Chern-Simons equations of motion again look homogenous, namely $G = 0$.
In terms of this shifted field the action $\exp(i S^\lambda(\hat a))$
from above, equivalently, reads
$$
  \exp(i S^\lambda(\hat G)) = 
  \exp(
    i \int_X \tfrac{1}{2}(\hat G \cup \hat G - (\tfrac{1}{2}\hat \lambda)^2)
  )
  \,.
$$
For the case $k = 1$, 
this is the form of the action functional for the 7d Chern-Simons dual 
of the 2-form gauge field on the M5-brane
first given as (3.6) in \cite{Witten96}

\medskip

In the language of twisted cohomological structures,
def. \ref{TwistedCStructures}, we may summarize this situation
as follows: 
{\it In order for the action functional of higher abelian Chern-Simons theory 
to be correctly divisible, the images of the fields in $\mathbb{Z}_2$-cohomology
need to form a \emph{twisted Wu-structure}, \cite{II}.Therefore the fields themselves
need to constitute a \emph{twisted $\lambda$}-structure. For $k = 1$ this is
a \emph{twisted String-structure} \cite{SSSIII} and explains the quantization
condition on the C-field in 11-dimensional supergravity. }

\vspace{3mm}
In \cite{HopkinsSinger} a formalization of the above situation
has been given in terms of a notion there called 
\emph{differential integral Wu structures}. In the following
section we explain how this follows from the notion of 
twisted Wu structures  \cite{II} with the twist taken in $\mathbb{Z}_2$-coefficients.
Then we refine this to a formalization to \emph{twisted differential Wu structures}
with the twist taken in smooth circle $n$-bundles.

%%%%%%%%%%%%%%%%%%%%%%%%%%%%%%%%%%%%%%%%%%%%%%%%%%%%%%%%%%%%%
\subsection{Twisted differential smooth Wu structures}
\label{DifferentialIntegralWuStructures}
%%%%%%%%%%%%%%%%%%%%%%%%%%%%%%%%%%%%%%%%%%%%%%%%%%%%%%%%%%%%%

We discuss some general aspects of smooth and differential refinements
of $\mathbb{Z}_2$-valued universal characteristic classes. For the special 
case of \emph{Wu classes} we show how these notions reduce to the 
definition of \emph{differential integral Wu structures} given in 
\cite{HopkinsSinger}. We then construct a refinement of these structures
that lifts the twist from $\mathbb{Z}_2$-valued cocycles to smooth circle $n$-bundles.
This further refinement of integral Wu structures 
is what underlies the model for the supergravity
C-field in section \ref{supergravityCField}.

\medskip

Recall from 
\cite{SSSIII, FiorenzaSatiSchreiberII}
%prop. \ref{SpinCAsHomotopyFiberProductOfU1AndSO} 
the
characterization of $\mathrm{Spin}^c$ as the 
loop space object of the homotopy pullback
$$
  \raisebox{20pt}{
  \xymatrix{
    \mathbf{B}\mathrm{Spin}^c \ar[r]
    \ar[d]	
	& 
	  \mathbf{B} U(1) \ar[d]^{\mathbf{c}_1\, \mathrm{mod}\, 2}
	\\
	\mathbf{B} \mathrm{SO}
	\ar[r]^{\mathbf{w}_2}
	&
	\mathbf{B}^2 \mathbb{Z}_2
  }}
  \;.
$$
For general $n \in \mathbb{N}$ the analog of the first Chern
class mod 2 appearing here is the higher Dixmier-Douady class
mod 2
$$
  \mathbf{DD}_{\mathrm{mod}\, 2}
  : 
  \xymatrix{
  \mathbf{B}^n U(1)
  \ar[r]^{\mathrm{DD}}
  &
  \mathbf{B}^{n+1} \mathbb{Z}
  \ar[rr]^{\mathrm{mod}\, 2}
  &&
  \mathbf{B}^{n+1} \mathbb{Z}_2
  }
  \,.
$$
Let now 
$$
  \nu_{n+1} : \mathbf{B} \mathrm{SO}  \to \mathbf{B}^{n+1} \mathbb{Z}_2
$$
be a representative of the universal \emph{Wu class} in degree $n+1$.

\vspace{3mm}
In the spirit of twisted structures in \cite{Wa, SSSIII, Sati10Twist, tw, II},
def. \ref{TwistedCStructures},
we have
\begin{definition}
  \label{IntegralUniversalWuClasses}
Let $\mathrm{Spin}^{\nu_{n+1}}$ be the loop space object of the 
homotopy pullback
$$
  \raisebox{20pt}{
  \xymatrix{
    \mathbf{B} \mathrm{Spin}^{\nu_{n+1}}
	\ar[r]
    \ar[d]^{\nnu_{n+1}^{\mathrm{int}}}
	& \mathbf{B}\mathrm{SO}
	\ar[d]^{\nu_{n+1}}
	\\
	\mathbf{B}^n U(1)
	 \ar[r]^{\hspace{-2mm}\mathrm{mod}\,2}
	 &
	\mathbf{B}^{n+1} \mathbb{Z}_2
  }
  }\;.
$$
We call the left vertical morphism $\nnu_{n+1}^{\rm int}$ appearing here
the \emph{universal smooth integral Wu structure} in degree $n+1$.
\end{definition}
A morphism of stacks
$$
  \nnu_{n+1} : X \to \mathbf{B} \mathrm{Spin}^{\nu_{n+1}}
$$
is a choice of orientation structure on $X$ together with a choice of
smooth integral Wu structure lifting the corresponding Wu class
$\nu_{n+1}$.
\begin{example}
  \label{FirstPontrjaginAsIntegralWu}
  The smooth first fractional Pontrjagin class $\tfrac{1}{2}\mathbf{p}_1$, 
  from prop. \ref{FirstFracPontryagin}, 
  fits into a diagram
  $$
  \raisebox{20pt}{
  \xymatrix{
    \mathbf{B} \mathrm{Spin}
	\ar@/^1pc/[rrd]
	\ar@/_1pc/[ddr]_{\tfrac{1}{2}\mathbf{p}_1}
	\ar@{-->}[dr]^{u}
    \\
    &\mathbf{B} \mathrm{Spin}^{\nu_{4}}
	\ar[r]
    \ar[d]^{\nnu_{4}^{\mathrm{int}}}
	& \mathbf{B}\mathrm{SO}
	\ar[d]^{\nu_{4}}
	\\
	& \mathbf{B}^3 U(1)
	 \ar[r]^{\hspace{-2mm}\mathrm{mod}\,2}
	 &
	\mathbf{B}^{4} \mathbb{Z}_2~~.
  }
  }
  $$
  In this sense we may think of $\tfrac{1}{2}\mathbf{p}_1$ as being the 
  integral and, moreover, smooth refinement of the universal degree-4 Wu class on 
  $\mathbf{B}\mathrm{Spin}$.
%\proof
  Using the defining property of $\tfrac{1}{2}\mathbf{p}_1$, this 
  follows with the results discussed in appendix E.1 of \cite{HopkinsSinger}.
\end{example}
%\endofproof
\begin{proposition}
  \label{RelationToHopkinsSingerDifferentialWu}
  Let $X$ be a smooth manifold equipped with orientation
  $$
    o_X : X \to \mathbf{B} \mathrm{SO}
  $$
  and consider its Wu-class $[\nu_{n+1}(o_X)] \in H^{n+1}(X, \mathbb{Z}_2)$
  $$
    \nu_{n+1}(o_X) : 
    \xymatrix{
	  X \ar[r]^{\hspace{-2mm}o_X} 
	  & 
	  \mathbf{B}\mathrm{SO}
	   \ar[r]^{\hspace{-3mm}\nu_{n+1}}
	  &
	  \mathbf{B}^{n+1}\mathbb{Z}_2
	}	
	\,.
  $$  
  The $n$-groupoid 
  $\hat {\mathbf{DD}}_{\mathrm{mod}2 }\mathrm{Struc}_{[\nu_{2k}]}(X)$
  of $[\nu_{n+1}]$-twisted differential 
  $\mathbf{DD}_{\mathrm{mod} 2}$-structures, according to def. \ref{TwistedCStructures},
  hence the homotopy pullback
  $$
    \raisebox{20pt}{
    \xymatrix{
	  \hat {\mathbf{DD}}_{\mathrm{mod}2 }\mathrm{Struc}_{[\nu_{n+1}]}(X)
	  \ar[rr]
	  \ar[d]
	  &&
	  {*}
	  \ar[d]^{\nu_{n+1}(o_X)}
	  \\
	  \mathbf{H}(X, \mathbf{B}^3 U(1)_{\mathrm{conn}})
	  \ar[rr]^{\hat {\mathbf{DD}}_{\mathrm{mod}\, 2}}
	  &&
	  \mathbf{H}(X, \mathbf{B}^{n+1} \mathbb{Z}_2)~~,
	}
	}
  $$
  categorifies the groupoid $\hat {\mathcal{H}}^{n+1}_{\nu_{n+1}}(X)$
  of \emph{differential integral Wu structures} as in def. 2.12 of \cite{HopkinsSinger}:
  its 1-truncation is equivalent to the groupoid defined there
  $$
    \tau_1 \hat {\mathbf{DD}}_{\mathrm{mod}2 }\mathrm{Struc}_{[\nu_{n+1}]}(X)
	\simeq
	\hat  {\mathcal{H}}^{n+1}_{\nu_{n+1}}(X)
	\,.
  $$
\end{proposition}
\proof
  By prop. \ref{BnU1conn}, the canonical presentation  
  of $\mathbf{DD}_{\mathrm{mod} 2}$
  via the Dold-Kan correspondence is given by an epimorphism of chain complexes of sheaves,
  hence by a fibration in $[\mathrm{CartSp}^{\mathrm{op}}, \mathrm{sSet}]_{\mathrm{proj}}$.
  Precisely,  the composite
  $$
    \hat {\mathbf{DD}}_{\mathrm{mod}\, 2}
	:
	\xymatrix{
	  \mathbf{B}^n U(1)_{\mathrm{conn}}
	  \ar[r]
	  &
	  \mathbf{B}^n U(1)
	  \ar[r]^{\mathrm{DD}}
	  &
	  \mathbf{B}^{n+1} \mathbb{Z}
	  \ar[rr]^{\mathrm{mod}\,2}
	  &&
	  \mathbf{B}^{n+1} \mathbb{Z}_2
	}
  $$
  is presented by the vertical sequence of morphisms of chain complexes
  $$
    \raisebox{30pt}{
    \xymatrix{
	  \mathbb{Z}~ \ar@{^{(}->}[r]
	  \ar[d]
	  &
	  C^\infty(-, \mathbb{R})~
	  \ar[r]^{~~d_{\mathrm{dR}} \mathrm{log}}~
	  \ar[d]
	  &
	  \Omega^1(-)
	  \ar[r]^{d_{\mathrm{dR}}}
	  \ar[d]
	  &
	  \cdots
	  \ar[r]^{\hspace{-2mm}d_{\mathrm{dR}}}
	  &
	  \Omega^n(-)
	  \ar[d]
	  \\
	  \mathbb{Z}~ \ar@{^{(}->}[r]
	  \ar[d]
	  &
	  C^\infty(-, \mathbb{R})
	  \ar[r]
	  \ar[d]
	  &
	  0
	  \ar[r]
	  \ar[d]
	  &
	  \cdots
	  \ar[r]
	  &
	  0
	  \ar[d]	  
	  \\
	  \mathbb{Z} \ar[r]
	  \ar[d]
	  &
	  0
	  \ar[r]
	  \ar[d]
	  &
	  0
	  \ar[r]
	  \ar[d]
	  &
	  \cdots
	  \ar[r]
	  &
	  0
	  \ar[d]	  
	  \\
	  \mathbb{Z}_2 \ar[r]
	  &
	  0
	  \ar[r]
	  &
	  0
	  \ar[r]
	  &
	  \cdots
	  \ar[r]
	  &
	  0  
	}
	}
  $$  
  We may therefore compute the defining 
  homotopy pullback for $\hat {\mathbf{DD}}_{\mathrm{mod}2 }\mathrm{Struc}_{[\nu_{n+1}]}(X)$
  as an ordinary fiber product of the corresponding simplicial sets of 
  cocycles. The claim then follows by inspection.
\endofproof
\begin{remark}
  Explicitly, a cocycle in 
  $
    \tau_1 \hat {\mathbf{DD}}_{\mathrm{mod}2 }\mathrm{Struc}_{[\nu_{n+1}]}(X)$
 is identified with a {\v C}ech cocycle with coefficients in
 the Deligne complex
 $$
  \left(
   \xymatrix{
	  \mathbb{Z} ~\ar@{^{(}->}[r]
	  &
	  C^\infty(-, \mathbb{R})~
	  \ar[r]^{~d_{\mathrm{dR}} \mathrm{log}}
	  &
	  ~
	  \Omega^1(-)
	  \ar[r]^{~~d_{\mathrm{dR}}}
	  &
	  \cdots
	  \ar[r]^{\hspace{-3mm}d_{\mathrm{dR}}}
	  &
	  \Omega^n(-)
   }
   \right)\;,
  $$
  such that the underlying $\mathbb{Z}[n+1]$-valued cocycle modulo 2
  equals the given cocycle for $\nu_{n+1}$. A coboundary between two such
  cocycles is a gauge equivalence class of ordinary {\v C}ech-Deligne cocycles
  such that their underlying $\mathbb{Z}$-cocycle vanishes modulo 2. 
  Cocycles of this form are precisely those that arise by multiplication with 
  2 or arbitrary {\v C}ech-Deligne cocycles. 
  This is the groupoid structure discussed on p. 14 of \cite{HopkinsSinger},
  there in terms of singular cohomology instead of {\v C}ech cohomology.
  \label{UnwindingDifferentialWuStructureClasses}
\end{remark}
We now consider another twisted differential structure, which 
refines these twisting integral Wu structures 
to \emph{smooth} integral Wu structures, of def. \ref{IntegralUniversalWuClasses}.
\begin{definition}
  For $n \in \mathbb{N}$, write $\mathbf{B}^n U(1)_{\mathrm{conn}}^{\nu_{n+1}}$
  for the homotopy pullback of smooth moduli $n$-stacks
  $$
    \raisebox{20pt}{
    \xymatrix{
	  \hat{\mathbf{Wu}}^{\nu_{n+1}}
	  \ar[rrr]
	  \ar[d]
	  &&&
	  \mathbf{B}^n U(1)_{\mathrm{conn}}
	  \ar[d]
	  \\
	  \mathbf{B} \mathrm{Spin}^{\nu_{n+1}}
		 \times
	   \mathbf{B}^n U(1)
	  \ar[rrr]^{~~~~\nnu_{n+1}^{\mathrm{int}} +  2 \mathbf{DD}}
	  &&&
	  \mathbf{B}^n U(1)~~~.
	}
	}
  $$
  where $\nnu_{n+1}^{\mathrm{int}}$ is the universal smooth integral Wu class from 
  def. \ref{IntegralUniversalWuClasses}, and where
  $2 \mathbf{DD} : \mathbf{B}^n U(1) \to \mathbf{B}^n U(1)$ is the canonical 
  smooth refinement of the operation of multiplication by 2 on integral cohomology.
  We call this the smooth moduli $n$-stack of \emph{smooth differential Wu structures}.
  \label{SmoothDifferentialWuStructures}
\end{definition}
By construction, a morphism $X \to \hat{\mathbf{Wu}}^{\nu_{n+1}}$
classifies also all possible orientation structures and 
smooth integral lifts of their Wu structures. In applications one typically
wants to fix an integral Wu structure lifting a given Wu class. 
This is naturally formalized by
the following construction.
\begin{definition}
  For $X$ an oriented manifold, and 
  $$
    \nnu_{n+1} : X \to \mathbf{B}\mathrm{Spin}^{\nu_{n+1}}
  $$
  a given smooth Wu structure, def. \ref{IntegralUniversalWuClasses},
  write
  $\mathbf{H}_{\nnu_{n+1}}(X, \mathbf{B}^n U(1)_{\mathrm{conn}})$
  for the $n$-groupoid of cocycles whose underlying 
  smooth integral Wu structure is $\nnu_{n+1}$, hence for the homotopy pullback
  $$
    \xymatrix{
	  \mathbf{H}_{\nnu_{n+1}}(X, \mathbf{B}^n U(1)_{\mathrm{conn}})
	  \ar[r]
	  \ar[d]
	  &
	  \mathbf{H}(X, \hat{\mathbf{Wu}}^{\nu_{n+1}})
	  \ar[d]
	  \\
	  \mathbf{H}(X, \mathbf{B}^n U(1))
	  \ar[r]^<<<<<<<{(\nnu_{n+1}, \mathrm{id})}
	  \ar[d]
	  & 
	  \mathbf{H}(X, \mathbf{B}\mathrm{Spin}^{\nu_{n+1}} \times \mathbf{B}^n U(1) )
	  \ar[d]
	  \\
	  {*} \ar[r]^{\hspace{-12mm}\nnu_{n+1}}& \mathbf{H}(X, \mathbf{B}\mathrm{Spin}^{\nu_{n+1}} )~~~.
	}
  $$
  \label{RestrictionOfDifferentialWuStructuresToFixedSmoothWuStructure}
\end{definition}
\begin{proposition}
 \label{RefinementOfDifferentialWuClassesTonBundles}
  Cohomology with coefficients in $\hat{\mathbf{Wu}}^{\nu_{n+1}}$
  over a given smooth integral Wu structure coincides with the corresponding
  differential integral Wu structures:
  $$
    \hat H_{\nu_{n+1}}^{n+1}(X) \simeq 
	H_{\nnu_{n+1}}(X, \mathbf{B}^n U(1)_{\mathrm{conn}})
	\,.
  $$
\end{proposition}
\proof
  Let $\check{C}(\mathcal{U})$ be the {\v C}ech-nerve of a good open cover $\mathcal{U}$ of $X$.
  By prop. \ref{BnU1conn} the canonical 
  presentation of $\mathbf{B}^n U(1)_{\mathrm{conn}} \to \mathbf{B}^n U(1)$ 
  is a projective fibration. Since $\check{C}(\mathcal{U})$ 
  is projectively cofibrant (it is a projectively cofibrant replacement of $X$) and $[\mathrm{CartSp}^{\mathrm{op}}, \mathrm{sSet}]_{\mathrm{proj}}$ 
  is a simplicial model category, the morphism of {\v{C}}ech cocycle simplicial sets
  $$
    [\mathrm{CartSp}^{\mathrm{op}}, \mathrm{sSet}](\check{C}(\mathcal{U}), \mathbf{B}^n U(1)_{\mathrm{conn}})
	\to
    [\mathrm{CartSp}^{\mathrm{op}}, \mathrm{sSet}](\check{C}(\mathcal{U}), \mathbf{B}^n U(1))
  $$
  is a Kan fibration.
  Hence, its
  homotopy pullback 
  may be computed as the ordinary pullback of simplicial sets of this map. 
  The claim then follows by inspection.
  
  Explicitly, in this presentation a cocycle in the pullback is
  a pair $\{a, \hat G\}$ of a cocycle $a$ for a circle $n$-bundle and a 
  Deligne cocycle $\hat G$ with underlying bare cocycle $G$, 
  such that there is an equality of degree-n
  {\v C}ech $U(1)$-cocycles
  $$
    G = \nnu_{n+1} + 2 a
	\,.
  $$
  A gauge transformation between two such cocycles is a pair of {\v C}ech cochains
  \{$\hat \gamma$, $\alpha$\} such that $\gamma = 2 \alpha$ (the cocycle 
  $\nnu_{n+1}$ being held fixed). 
  This means that the gauge transformations acting on 
  a given $\hat G$ solving the above constraint
  are precisely all the Deligne cochains, but multiplied by 2.
  This is again the explicit description of $\hat H_{\nu_{n+1}}(X)$
  from remark \ref{UnwindingDifferentialWuStructureClasses}.
  \endofproof

%%%%%%%%%%%%%%%%%%%%%%%%%%%%%%%%%%%%%%%%%%%%%%%%%%%%%%%%%%%%%%%%%%%%%
\section{The C-field}
\label{supergravityCField}
\index{supergravity!C-field}
%%%%%%%%%%%%%%%%%%%%%%%%%%%%%%%%%%%%%%%%%%%%%%%%%%%%%%%%%%%%%%%%%%%%%
In this section we describe our model for the C-field, first for bulk fields, and then 
for fields in the presence of boundaries and/or M5-branes.

%%%%%%%%%%%%%%%%%%%%%%%%%%%%%%%%%%%%%%%%%%%%%%%%%%%%%%%%%%%%%
\subsection{The moduli 3-stack of the C-field}
\label{TheModuli3StackOfTheCField}
%%%%%%%%%%%%%%%%%%%%%%%%%%%%%%%%%%%%%%%%%%%%%%%%%%%%%%%%%%%%%

As we have reviewed above in 
section \ref{CSWithBackgroundCharge}, 
the flux quantization condition for the C-field
derived in \cite{WittenFluxQuantization} is the equation
\(
\label{eq.congruent.mod.2}
  [G_4] = \tfrac{1}{2}p_1 ~{\rm mod}~ 2
  \;\;\;
 \text{in}~~  H^4(X, \mathbb{Z})
\)
in integral cohomology,  
where $[G_4]$ is the cohomology class 
%of the field strength 
of the C-field itself, and $\tfrac{1}{2}p_1$ is the first fractional Pontrjagin 
class of the Spin manifold $X$. One can equivalently rewrite \eqref{eq.congruent.mod.2} as
\(
\label{eq.integral}
  [G_4] = \tfrac{1}{2}p_1+2a
  \;\;\;
 \text{in}~~ H^4(X, \mathbb{Z})\;,
\)
where $a$ is some degree 4 integral cohomology class on $X$. 
By the discussion in section \ref{DifferentialIntegralWuStructures}, 
the correct formalization of this for \emph{fixed} Spin 
structure \footnote{The dependence of the partition function of the C-field on 
the Spin structure(s) is discussed in \cite{Spinc}.}
is to regard the gauge equivalence class of the 
C-field as a differential integral Wu class 
relative to the integral Wu class $\nu_4^{\mathrm{int}} = \tfrac{1}{2}p_1$,
example \ref{FirstPontrjaginAsIntegralWu}, of that Spin structure. 
By prop. \ref{RefinementOfDifferentialWuClassesTonBundles} 
and prop. \ref{FirstFractionalDifferentialPontrjagin}, the natural
refinement of this to a smooth moduli 3-stack of C-field configurations
and arbitrary spin connections is the homotopy pullback of  smooth 3-stacks
$$
  \raisebox{20pt}{
  \xymatrix{
    \hat{\mathbf{Wu}}^{\nu_{n+1}}
	\ar[rr]
	\ar[d]
	&& \mathbf{B}^3 U(1)_{\mathrm{conn}}
	\ar[d]
    \\
    \mathbf{B} \mathrm{Spin}_{\mathrm{conn}}
	\times 
	\mathbf{B}^3 U(1)
	\ar[rr]^<<<<<<<<<{~~\tfrac{1}{2}\hat {\mathbf{p}}_1 + 2 \mathbf{DD}}
	&&
	\mathbf{B}^3 U(1)~~.
  }
  }
$$
Here the moduli stack in the bottom left is that of the field of gravity
(spin connections) together with an auxiliary circle 3-bundle / 2-gerbe.
Following the arguments in \cite{FiorenzaSatiSchreiberII} 
(the traditional ones as well as the new ones presented there),
we take this auxiliary circle 3-bundle to be 
the Chern-Simons circle 3-bundle of an $E_8$-principal bundle.
According to prop. \ref{aDifferentially} this is formalized 
on smooth higher moduli stacks by further pulling back along the smooth refinement
$$
  \mathbf{a} : \mathbf{B}E_8 \to \mathbf{B}^3 U(1)
$$
of the canonical universal 4-class $[a] \in H^4(B E_8, \mathbb{Z})$.
Therefore, we are led to formalize the
\emph{$E_8$-model for the C-field} as follows. 
\begin{definition}
The \emph{smooth moduli 3-stack of Spin connections and C-field configurations} 
in the $E_8$-model is the homotopy pullback $\mathbf{CField}$ of the moduli
$n$-stack of smooth differential Wu structures $\mathbf{B}^n U(1)_{\mathrm{conn}}^{\nu_4}$,
def. \ref{SmoothDifferentialWuStructures},
to Spin connections and $E_8$-instanton configurations, hence the homotopy pullback
$$
  \raisebox{20pt}{
  \xymatrix{
    \mathbf{CField} \ar[rr]^{} \ar[d] 
	 &&  \hat{\mathbf{Wu}}^{\nu_4}
	 \ar[d]
    \\
   \mathbf{B}\mathrm{Spin}_{\mathrm{conn}}
	  \times
	  \mathbf{B}{E_8}
    \ar[rr]^{\hspace{-2mm}
	  (u, \mathbf{a})
	}
	&&
   \mathbf{B}\mathrm{Spin}^{\nu_4} \times \mathbf{B}^3 U(1)~~,
	}
	}
$$
where $u$ is the canonical morphism from example \ref{FirstPontrjaginAsIntegralWu}.
\label{CField}
\end{definition}
\begin{remark}
By the pasting law, prop. \ref{PastingLawForPullbacks},
$\mathbf{CField}$ is equivalently given as the homotopy pullback
$$
  \raisebox{20pt}{
  \xymatrix{
    \mathbf{CField} \ar[rrr]^{\hat{\mathbf{G}}_4} \ar[d] 
	 &&&  \mathbf{B}^3 U(1)_{\mathrm{conn}}\ar[d]
    \\
   \mathbf{B}\mathrm{Spin}_{\mathrm{conn}}
	  \times
	  \mathbf{B}{E_8}
    \ar[rrr]^{
	  \frac{1}{2}\mathbf{p}_1 + 2 \mathbf{a}
	}
	&&&
   \mathbf{B}^3 U(1)~~.
	}
	}
$$
Spelling out this definition, a C-field configuration 
$$
  (\nabla_{\mathfrak{so}}, \nabla_{b^2 \mathbb{R}}, P_{E_8})
  :
  X \to \mathbf{CField}
$$
on a smooth manifold $X$  is the datum of 
{\it 
\begin{enumerate}
 \item a principal $\mathrm{Spin}$-bundle with $\mathfrak{so}$-connection $(P_{\mathrm{Spin}},\nabla_{\mathfrak{so}})$ on $X$;
\item a principal $E_8$-bundle $P_{E_8}$  on $X$;
\item a $U(1)$-2-gerbe with connection $(P_{\mathbf{B}^2U(1)},\nabla_{\mathbf{B}^2U(1)})$ on $X$;
\item a choice of equivalence of $U(1)$-2-gerbes between  
  between $P_{\mathbf{B}^2U(1)}$ and the image of 
  $P_{\mathrm{Spin}}\times_XP_{E_8}$ via $\frac{1}{2}\mathbf{p}_1 + 2 \mathbf{a}$.
\end{enumerate}
}
\label{CFieldAs3Bundle}
\end{remark}
\noindent It is useful to observe that there is the following further equivalent reformulation of this 
definition.
\begin{proposition}
  The moduli 3-stack $\mathbf{CField}$ from def. \ref{CField} is equivalently
  the homotopy pullback
$$
  \raisebox{20pt}{
  \xymatrix{
    \mathbf{CField} \ar[rrr] \ar[d] 
	 &&& \Omega^4_{\mathrm{cl}}\ar[d]
    \\
   \mathbf{B}\mathrm{Spin}_{\mathrm{conn}}
	  \times
	  \mathbf{B}{E_8}
    \ar[rrr]^{~~
	  (\frac{1}{2}\mathbf{p}_1 + 2 \mathbf{a})_{\mathrm{dR}}
	}
	&&&
  \mathbf{\flat}_{\mathrm{dR}} \mathbf{B}^4 \mathbb{R}~~,
	  }
	  }
$$
where, by remark. \ref{DeRhamImagesOfDifferentialRefinements}, 
the bottom morphism of higher stacks is presented by the
correspondence of simplicial presheaves
$$
  \raisebox{20pt}{
  \xymatrix{
    \mathbf{B}\mathrm{Spin}_{\mathrm{conn}}
	\times
	(\mathbf{B}E_{8})_{\mathrm{diff}}
	\ar[r]
    \ar@{->>}[d]^\wr
      &
    \mathbf{B}\mathrm{Spin}_{\mathrm{diff}}
	\times
	(\mathbf{B}E_{8})_{\mathrm{diff}}
	  \ar[rr]^<<<<<<<{~~~~(\tfrac{1}{2}\mathbf{p}_1 + 2 \mathbf{a})_{\mathrm{diff}}} 
    \ar@{->>}[d]^\wr	
	  && 
	  \mathbf{B}^3 U(1)_{\mathrm{diff}}
    \ar[r]^{\mathrm{curv}}
    \ar@{->>}[d]^\wr
    &
    \mathbf{\flat}_{\mathrm{dR}} \mathbf{B}^{4}\mathbb{R} 
    \\
   \mathbf{B}\mathrm{Spin}_{\mathrm{conn}}\times
	  \mathbf{B}{E_8} 
	  \ar[r] 
	  &
	  \mathbf{B}\mathrm{Spin}\times \mathbf{B}{E_8}
	  \ar[rr]^{~~\tfrac{1}{2}\mathbf{p}_1 + 2 \mathbf{a}} 
	  && 
	  \mathbf{B}^3 U(1)
  }
  }
  \;.
$$
Moreover, it is equivalently the homtopy pullback
$$
  \raisebox{20pt}{
  \xymatrix{
    \mathbf{CField} \ar[rrr] \ar[d] 
	 &&& \Omega^4_{\mathrm{cl}}\ar[d]
    \\
   \mathbf{B}\mathrm{Spin}_{\mathrm{conn}}
	  \times
	  \mathbf{B}{E_8}
    \ar[rrr]^{~~
	  (\frac{1}{4}\mathbf{p}_1 + \mathbf{a})_{\mathrm{dR}}
	}
	&&&
  \mathbf{\flat}_{\mathrm{dR}} \mathbf{B}^4 \mathbb{R}~~,
	  }
	  }
$$
where now the bottom morphism is the composite of the bottom morphism before,
postcomposed with the morphism
$$
  \tfrac{1}{2} 
    : 
   \mathbf{\flat}_{\mathrm{dR}} \mathbf{B}^4 \mathbb{R}
     \to 
   \mathbf{\flat}_{\mathrm{dR}} \mathbf{B}^4 \mathbb{R}
$$
that is given, via Dold-Kan, by division of differential forms by 2.
\end{proposition}
\proof
 By the pasting law 
 for homotopy pullbacks, prop. \ref{PastingLawForPullbacks}, 
 the first homotopy pullback above may be computed as
 two consecutive homotopy pullbacks
 $$
  \raisebox{20pt}{
  \xymatrix{
    \mathbf{CField} \ar[rr] \ar[d] 
	&&
	\mathbf{B}^n U(1)_{\mathrm{conn}}
	\ar[r]
	\ar[d]
	 & 
	 \Omega^4_{\mathrm{cl}}\ar[d]
    \\
   \mathbf{B}\mathrm{Spin}_{\mathrm{conn}}
	  \times
	  \mathbf{B}{E_8}
    \ar[rr]^{~~~~
	  \tfrac{1}{2} {\mathbf{p}}_1 + 2  {\mathbf{a}}
	}
	&&
	\mathbf{B}^3 U(1)
	\ar[r]^{\hspace{-2mm}\mathrm{curv}}
	&
    \mathbf{\flat}_{\mathrm{dR}} \mathbf{B}^4 \mathbb{R}~~,
	  }
	  }
 $$
 which exhibits on the right the defining pullback of def. \ref{BnU1conn},
 and thus on the left the one from def. \ref{CField}.
 The statement about the second homotopy pullback above 
 follows analogously after noticing that
 $$
  \xymatrix{
     \Omega^4_{\mathrm{cl}}\ar[rrr]^{\tfrac{1}{2}} \ar[d] 
	 &&& \Omega^4_{\mathrm{cl}}\ar[d]
    \\
     \mathbf{\flat}_{\mathrm{dR}} \mathbf{B}^4 \mathbb{R}    \ar[rrr]^{
	  \tfrac{1}{2}}
	&&&
  \mathbf{\flat}_{\mathrm{dR}} \mathbf{B}^4 \mathbb{R}
  }
$$
is a homotopy pullback.
\endofproof

\noindent It is therefore useful to introduce labels as follows.
\begin{definition}
We label the structure morphism of the above composite homotopy pullback as
$$
  \raisebox{20pt}{
  \xymatrix{
    \mathbf{CField} 
	 \ar[rr]^{\hspace{-2mm}\hat G_4}_>{\ }="s1" 
	 \ar[d]
	 && 
    \mathbf{B}^3 U(1)_{\mathrm{conn}}
     \ar[r]^{~~~~\mathcal{G}_4}_>{\ }="s2" 
	  \ar[d]^{G_4}
	  &
     \Omega^4_{\mathrm{cl}} 
	 \ar[d]
   \\
	 \mathbf{B}\mathrm{Spin}_{\mathrm{conn}}
	 \times
	 \mathbf{B}{E_8}
   \ar[rr]_<<<<<<<<<{~~\frac{1}{2}\mathbf{p}_2 + 2 \mathbf{a}}^<{\ }="t1"
    &&
   \mathbf{B}^3 U(1)
    \ar[r]_<<<<<<{\mathrm{curv}}^<{\ }="t2"
	&
   \mathbf{\flat}_{\mathrm{dR}} \mathbf{B}^4 U(1)~~.
   \ar@{=>}_{H_3}^\simeq "s1"; "t1"
   \ar@{=>}^\simeq "s2"; "t2"
  }
  }
$$
Here $\hat G_4$ sends a C-field configuration to an underlying circle 3-bundle
with connection, whose curvature 4-form is $\mathcal{G}_4$.
\end{definition}
\begin{remark}
 \label{CommentOnTheTwoIncarnationsOfDefOfCField}
  These equivalent reformulations 
  point to two statements.
  %show two things.
  \begin{enumerate}
    \item 
	  The C-field model may be thought of as containing
	  $E_8$-\emph{pseudo-connections}, remark \ref{DeRhamImagesOfDifferentialRefinements}. 
	  That is, there is a higher gauge in which
	  a field configuration consists of an $E_8$-connection on an $E_8$-bundle
	   -- even though there is no dynamical $E_8$-gauge field in 11d supergravity --
	  but where gauge transformations are allowed to freely shift these connections.
    \item
	  There is a precise sense in which imposing 
	  the quantization condition (\ref{eq.integral})
	  on integral cohomology is equivalent to imposing the condition
	  $[G_4] = \tfrac{1}{4}p_1 + a$ in de Rham cohomology / real singular cohomology.
  \end{enumerate}
\end{remark}
\begin{observation}
  When restricted to a fixed Spin connection, gauge equivalence classes 
  of configurations classified by $\mathbf{CField}$ naturally form a torsor
  over the ordinary degree-4 differential cohomology $H^4_{\mathrm{diff}}(X)$.
  \label{RestrictionToFixedSpinConnection}
\end{observation}
\proof
  By the general discussion of differential integral Wu-structures
  in section \ref{DifferentialIntegralWuStructures}.
\endofproof

\noindent We now comment on the relation to the proposal in \cite{DFM}.
\begin{remark}
  The first item in remark \ref{CommentOnTheTwoIncarnationsOfDefOfCField}
  finds its correspondence in equation (3.13) in \cite{DFM},
  where a definition of gauge transformation of the C-field  is proposed.
  The second item finds its correspondence in equation (3.26) there,
  where another model for the groupoid of C-field configurations is proposed.
  However, the immediate  
  translation of equation (3.25) used there, in the language
  of homotopy pullbacks 
  is given by the homotopy limit over the diagram
  $$
    \xymatrix{
	  &&&&
	  {*}
	  \ar[d]^0
	  \\
	  {*}
	  \ar[rr]^<<<<<<<{~~~~~~\tfrac{1}{2}\hat {\mathbf{p}}_1(g)}
	  \ar@/_1.2pc/[rrrr]_{\hat W_5}
	  &&
	  \mathbf{H}(X, \mathbf{B}^3 U(1)_{\mathrm{conn}})
	  \ar[r]^<<<<<{\mathrm{curv}}
	  &
	  \Omega^4_{\mathrm{cl}}(X)
	  \ar[r]
	  &
	  \mathbf{H}(X, \mathbf{B}^4 U(1)_{\mathrm{conn}})~~.
	}
  $$
  On gauge equivalence classes this becomes a torsor over 
  $H^4_{\mathrm{diff}}(X)$. So, by prop. \ref{RestrictionToFixedSpinConnection},
  for a fixed Spin connection this is equivalent to the model that we present here
  (which is naturally equivalent to the group of differential integral Wu structures),
  since any two torsors over a given group are equivalent.
  However, the equivalence is non-canonical, in general. 
  More precisely, for structures parameterized over spaces as here, the
  equivalence is in general non-natural, in the technical sense. 
\end{remark}

%%%%%%%%%%%%%%%%%%%%%%%%%%%%%%%%%%%%%%%%%%%%%%%%%%%%%%
\subsection{The homotopy type of the moduli 3-stack}
\label{TheHomotopyTypeOfTheCFieldModuliStack}
%%%%%%%%%%%%%%%%%%%%%%%%%%%%%%%%%%%%%%%%%%%%%%%%%%%%%%

We discuss now the homotopy type of the the 3-groupoid 
$$
  \mathbf{CField}(X)
  :=
  \mathbf{H}(X, \mathbf{CField})
$$
of C-field configurations over a given spacetime manifold $X$. 
In terms of gauge theory, 
its 0-th homotopy group is the set of \emph{gauge equivalence classes}
of field configurations, its first homotopy group is the 
set of \emph{gauge-of-gauge equivalence classes} of 
auto-gauge transformations of a given configuration, and so on.

\medskip

\begin{definition}
For $X$ a smooth manifold, 
let 
$$
  \raisebox{20pt}{
  \xymatrix{
 	& \mathbf{B}\mathrm{Spin}_{\mathrm{conn}}
    \ar[d]
	\\
    X 
	 \ar[r]^{P_{\mathrm{Spin}}}
	 \ar[ur]^{\nabla_{\mathfrak{so}}}
	 &
	\mathbf{B}\mathrm{Spin}
  }
  }
$$
be a fixed Spin structure with fixed Spin connection. 
The restriction of $\mathbf{CField}(X)$ to this fixed Spin connection is the
homotopy pullback
$$
  \raisebox{20pt}{
  \xymatrix{
    \mathbf{CField}(X)_{P_{\mathrm{Spin}}}
	\ar[rrr]
	\ar[d]
	&&&
	\mathbf{CField}(X)
	\ar[d]
	\\
	\mathbf{H}(X, \mathbf{B}E_8)
	\ar[rrr]^<<<<<<<<<<<<<<<{~~~((P_{\mathrm{Spin}},\nabla_{\mathfrak{so}}), \mathrm{id})}
	&&&
	\mathbf{H}(X, \mathbf{B}\mathrm{Spin}_{\mathrm{conn}} \times\mathbf{B}E_8)\;.
  }
  }
$$
\end{definition}
\begin{proposition}
  The gauge equivalence classes of $\mathbf{CField}(X)_{P_{\mathrm{Spin}}}$
  naturally surjects onto the differential integral Wu structures on $X$, relative to 
  $\tfrac{1}{2}p_1(P_{\mathrm{Spin}})\,\mathrm{mod}\,2$,
  (example \ref{FirstPontrjaginAsIntegralWu}):
  $$
    \xymatrix{
      \pi_0 \mathbf{CField}(X)_{P_{\mathrm{Spin}}}
	  \ar@{->>}[r]
	  &
	  \hat H^{n+1}_{\tfrac{1}{2}p_1(P_{\mathrm{Spin}})}(X)
	}
	\,.
  $$
  The gauge-of-gauge equivalence classes of the auto-gauge transformation
  of the trivial C-field configuration naturally surject onto 
  the singular cohomology
  $H^2_{\mathrm{sing}}(X, U(1))$ (see example \ref{RelationSingularSmoothCohomology}):
  $$
    \xymatrix{
      \pi_1 \mathbf{CField}(X)_{P_{\mathrm{Spin}}}
	  \ar@{->>}[r]
	  &
	  H^2_{\mathrm{sing}}(X, U(1))
	}    
	\,.
  $$
\end{proposition}
\proof
By def. \ref{CField} and the pasting law, prop. \ref{PastingLawForPullbacks},
we have a pasting diagram of homotopy pullbacks of the form
$$
  \raisebox{18pt}{
  \hspace{-5mm}
  \xymatrix@C=5pt{
    \mathbf{CField}(X)_{P_{\mathrm{Spin}}}
	\ar@{->>}[rr]
	\ar[d]
	&&
	\mathbf{H}_{\tfrac{1}{2}\mathbf{p}_1(P_{\mathrm{Spin}})}(X, 
	\mathbf{B}^n U(1)_{\mathrm{conn}})
	\ar[rrr]
	\ar[d]
	&&&
	\mathbf{H}(X,
  	  \hat{\mathbf{Wu}}^{\nu_4}
	)
	\ar[d]
    \\
    \mathbf{H}(X, \mathbf{B}E_8)
	\ar@{->>}[rr]^{\hspace{-3mm}\mathbf{H}(X, \mathbf{a})}
	&&
	\mathbf{H}(X, \mathbf{B}^3 U(1))
	\ar[r]^<<<<<<<<{\hspace{-3mm}(\nabla_{\mathfrak{so}}, \mathrm{id})}
	&
	\mathbf{H}(X, \mathbf{B}\mathrm{Spin}_{\mathrm{conn}} \times \mathbf{B}^3 U(1))
	\ar[rr]^{(u, \mathrm{id})}
	&&
	\mathbf{H}(X, \mathbf{B}\mathrm{Spin}^{\nu_4} \times \mathbf{B}^3 U(1))\;,
  }
  }
$$
where in the middle of the top row we identified, by 
def. \ref{RestrictionOfDifferentialWuStructuresToFixedSmoothWuStructure}, the
$n$-groupoid of smooth differential Wu structures lifting the smooth
Wu structure $\tfrac{1}{2}\mathbf{p}_1(P_{\mathrm{Spin}})$.
Due to prop. \ref{RefinementOfDifferentialWuClassesTonBundles}
we are, therefore, reduced to showing that the top left morphism is 
surjective on $\pi_0$.
But the bottom left morphism is surjective on $\pi_0$, by 
prop. \ref{aIsMaxCompactSubgroup}. Now, the morphisms surjective on
$\pi_0$ are precisely the \emph{effective epimorphisms} in 
$\infty \mathrm{Grpd}$,
and these are stable under pullback. Hence the first claim follows.

\vspace{2mm}
For the second, we use that
$$
  \pi_1 \mathbf{CField}(X)_{P_{\mathrm{Spin}}}
  \simeq
   \pi_0 \Omega \mathbf{CField}(X)_{P_{\mathrm{Spin}}}
$$
and that forming loop space objects (being itself a homotopy pullback)
commutes with homotopy pullbacks and with 
taking cocycles with coefficients in higher stacks, $\mathbf{H}(X,-)$.
Therefore, the image of the left square in the above under $\Omega$ is the
homotopy pullback
$$
  \raisebox{20pt}{
  \xymatrix{
    \Omega \mathbf{CField}(X)_{P_{\mathrm{Spin}}}
	\ar@{->>}[rr]
	\ar[d]
	&&
	\mathbf{H}_{\tfrac{1}{2}\mathbf{p}_1(P_{\mathrm{Spin}})}(X, 
	\hat{\mathbf{Wu}}^{\nu_4})
	\ar[d]
	\\
	C^\infty(X, E_8)
	\ar@{->>}[rr]^{\hspace{-3mm}\mathbf{H}(X, \Omega \mathbf{a})}
	&&
	\mathbf{H}(X, \mathbf{B}^2 U(1))~~~,
  }
  }
$$
where in the bottom left corner we used
$$
  \begin{aligned}
  \Omega \mathbf{H}(X, \mathbf{B}E_8)
  &\simeq
  \mathbf{H}(X, \Omega \mathbf{B}E_8)
  \\
  &\simeq
  \mathbf{H}(X, E_8)
  \\
  &\simeq
  C^\infty(X, E_8)~~,
  \end{aligned}
$$
and similarly for the bottom right corner. 
This identifies the bottom morphism on connected components as the
morphism that sends a smooth function $X \to E_8$ to its homotopy class
under the homotopy equivalence $E_8 \simeq_{15} B^2 U(1) \simeq K(\mathbb{Z},3)$,
which holds over the 11-dimensional $X$.
Therefore the bottom morphism is again surjective on $\pi_0$, and so is
the top morphism. The claim then follows with prop. \ref{RelationToHopkinsSingerDifferentialWu}.
\endofproof

%%%%%%%%%%%%%%%%%%%%%%%%%%%%%%%%%%%%%%%%%%%%%%%%%%%%%%%%
\subsection{The boundary moduli}
\label{CFieldRestrictionToTheBoundary}
%%%%%%%%%%%%%%%%%%%%%%%%%%%%%%%%%%%%%%%%%%%%%%%%%%%%%%%%

We now consider the moduli 3-stack of the C-field in the presence of a boundary
(possibly with more than one component). 
We will consider two variants, corresponding to two different boundary conditions
on the C-fiels:
The first,  $\mathbf{CField}^{\mathrm{bdr}}$ corresponds to the case when the 
field strength $G_4$ of the C-field vanishes on the boundary as a differential cocycle.
The second, $\mathbf{CField}^{\mathrm{bdr'}}$,
corresponds to the case when $G_4$ is zero as a cohomology class. 
Extensive discussion of boundary conditions can be found in \cite{Coh}.
See also the discussion in the companion article \cite{FiorenzaSatiSchreiberII}.

\vspace{3mm}
Let $\partial X$ be (a neighbourhood of) the boundary of the spacetime manifold $X$. 
The condition on the boundary configurations of the supergravity fields are
\begin{enumerate}
  \item 
    The C-field vanishes on the boundary, as a differential cocycle
	(in the Ho{\v r}ava-Witten model \cite{HoravaWitten} this follows by arguments
	as recalled for instance in section 3.1 of \cite{Falkowski}) or as a cohomology class;
  \item
    the $E_8$ bundle becomes equipped with a connection over the boundary, and hence
    becomes dynamical there. 
\end{enumerate}
We present now a natural morphism of 3-stacks
$$
  \mathbf{CField}^{\mathrm{bdr}} \to \mathbf{CField}
$$
into the moduli stack of bulk C-fields, def. \ref{CField}, 
such that C-field configurations on $X$ with the above mentioned behavior (with the strict condition $\hat{G}_4=0$) 
over $\partial X$
correspond to the \emph{relative twisted cohomology}, def.
\ref{TwistedCStructures}, with coefficients in this morphism, i.e., to
commuting diagrams of the form
$$
  \raisebox{20pt}{
  \xymatrix{
    \partial X_{}\ar[r]^<<<<<{\phi_{\mathrm{bdr}}} 
	\ar@{^{(}->}[d]
	  &  \mathbf{CField}^{\mathrm{bdr}} \ar[d]	  
	\\
	X \ar[r]^<<<<<<{\phi} &  \mathbf{CField}~~\;.
  }
  }
$$
\begin{definition}
Let
$$
  i 
    : 
  \mathbf{B}(\mathrm{Spin} \times E_8)_{\mathrm{conn}}
    \to
  \mathbf{CField}
$$
be the canonical morphism induced from the commuting
diagram of def. \ref{SmoothAndDifferentialRefinement},
for the differential characteristic maps 
prop. \ref{FirstFractionalDifferentialPontrjagin} and prop. \ref{aDifferentially}
and the universal property of the homotopy pullback defining $\mathbf{CField}$:
$$
  \raisebox{30pt}{
  \xymatrix{    
    \mathbf{B}(\mathrm{Spin} \times E_8)_{\mathrm{conn}}
	\ar@{-->}[dr]^i
	\ar[ddr]
	\ar@/^1pc/[drrrr]^{\frac{1}{2}\hat {\mathbf{p}}_1 + 2 \hat {\mathbf{a}}}
    \\
    &  \mathbf{CField}
	\ar[d]
	\ar[rrr]^{\hspace{-8mm}\hat{\mathbf{G}}_4}
	&&&
	\mathbf{B}^3 U(1)_{\mathrm{conn}}
	\ar[d]
    \\
    & \mathbf{B}\mathrm{Spin}_{\mathrm{conn}}
	\times
	\mathbf{B} E_8
	\ar[r]
	&
    \mathbf{B}\mathrm{Spin}
	\times
	\mathbf{B} E_8
	\ar[rr]^{\frac{1}{2} {\mathbf{p}}_1 + 2  {\mathbf{a}}}
	&&
	\mathbf{B}^3 U(1)~~\;.
  }
  }
$$
\end{definition}
\begin{remark}
The dashed morphism $i$ implements the condition that 
the $E_8$-bundle picks up a connection -- a dynamical gauge field -- on the boundary. 
Therefore, to naturally 
define $\mathbf{CField}^{\mathrm{bdr}}$ it only remains to model the condition 
that $\hat G_4$
vanishes on the boundary, either as a differential cocycle or in 
the underlying integral cohomology.  These requirements are immediately realized by further pulling 
back along the sequence of morphisms $* \xrightarrow{0} \Omega^{1\leq\bullet\leq 3}\to \mathbf{B}^3 U(1)_{\mathrm{conn}}$.
\end{remark}
\begin{definition}
  Let the moduli 3-stacks $\mathbf{CField}^{\mathrm{bdr}}$ and 
  $\mathbf{CField}^{\mathrm{bdr}'}$
  be defined as the consecutive homotopy pullbacks in this diagram
$$
  \raisebox{30pt}{
  \xymatrix{    
   \mathbf{CField}^{\mathrm{bdr}}\ar[d]\ar@/^1pc/[drrrr]\\
  \mathbf{CField}^{\mathrm{bdr}'}\ar[d]\ar@/^1pc/[drrrr]&&&&{*}\ar[d] \\
    \mathbf{B}(\mathrm{Spin} \times E_8)_{\mathrm{conn}}
	\ar@{-->}[dr]
	\ar[ddr]
	\ar@/^1pc/[drrrr]^{\frac{1}{2}\hat {\mathbf{p}}_1 + 2 \hat {\mathbf{a}}}
   &&&&{\Omega^{1\leq\bullet\leq 3}}\ar[d] \\
    &  \mathbf{CField}
	\ar[d]
	\ar[rrr]^{\hspace{-8mm}\hat{\mathbf{G}}_4}
	&&&
	\mathbf{B}^3 U(1)_{\mathrm{conn}}
	\ar[d]
    \\
    & \mathbf{B}\mathrm{Spin}_{\mathrm{conn}}
	\times
	\mathbf{B} E_8
	\ar[r]
	&
    \mathbf{B}\mathrm{Spin}
	\times
	\mathbf{B} E_8
	\ar[rr]^{\frac{1}{2} {\mathbf{p}}_1 + 2  {\mathbf{a}}}
	&&
	\mathbf{B}^3 U(1)~~\;.
  }
  }
$$
\end{definition}
A straightforward application of the  pasting law, prop. \ref{PastingLawForPullbacks}
and inspection of the definitions then gives
\begin{proposition}
  We have natural equivalences
  $$
    \mathbf{CField}^{\mathrm{bdr}} 
	  \simeq
	\mathrm{String}^{-2\hat{\mathbf{a}}}_{\mathrm{conn}}
	\,,
  $$
  and 
   $$
    \mathbf{CField}^{\mathrm{bdr}'} 
	  \simeq
	\mathrm{String}^{-2\hat{\mathbf{a}}}_{\mathrm{conn}'}
  $$
  of the moduli 3-stack of boundary C-field configurations, 
  with that of (nonabelian) $\mathrm{String}^{2\mathbf{a}}$ 2-connections,
  strict or weak, respectively, according to 
  \cite{FiorenzaSatiSchreiberII}.
  \label{BoundaryConfigsAreString2aConnections}
\end{proposition}
\begin{remark}
In \cite{FSS} we have given a detailed construction of these
2-stacks in terms of explicit differential form data.
In \cite{SSSIII} we have shown that these are the moduli 
2-stacks for heterotic background fields that 
satisfy the Green-Schwarz anomaly cancellation condition.
\end{remark}

%%%%%%%%%%%%%%%%%%%%%%%%%%%%%%%%%%%%%%%%%%%%%%%%%%%%%%%%%%%%%%%%%%%%%%
\subsection{Ho{\v r}ava-Witten boundaries and higher orientifolds}
%%%%%%%%%%%%%%%%%%%%%%%%%%%%%%%%%%%%%%%%%%%%%%%%%%%%%%%%%%%%%%%%%%%%%%

We now discuss a natural formulation of the origin of the 
Ho{\v r}ava-Witten boundary conditions \cite{HoravaWitten} in terms of 
higher stacks and nonabelian differential cohomology,
specifically, in terms of what we call \emph{membrane orientifolds}.
From this we obtain a corresponding refinement of the 
moduli 3-stack of C-field configurations which now explicitly 
contains the twisted $\mathbb{Z}_2$-equivariance of the
Ho{\v r}ava-Witten background.

\medskip

Earlier, around prop. \ref{PresentationOfUniversalCurvature} and
prop. \ref{DoldKanOnDeligne}, we invoked the Dold-Kan correspondence
in order to construct a higher stack from a chain complex of sheaves
of abelian groups. Now, in order to add a $\mathbb{Z}_2$-twist
to ordinary differential cohomology, we invoke 
the following nonabelian generalization of the
Dold-Kan correspondence. 
The discrete ingredients for that 
construction are discussed in some detail in \cite{BHS}. 
As a presentation of smooth higher stacks this is discussed
in section 2.1.7 of \cite{survey} and the concrete application
to higher orientifolds is in 4.4.3 there.

\medskip

\begin{definition}
  A \emph{crossed complex of groups}  $G_\bullet^\rho$ is 
  a complex of groups of the form
  $$
    \xymatrix{
      \cdots 
	  \ar[r]^{\delta_2}
	  &
	  G_2 
	  \ar[r]^{\delta_1}
	  &
	  G_1 
	  \ar[r]^{\delta_0}
	  &
	  G_0
	}
  $$
  with $G_{k \geq 2}$ abelian (but $G_1$ and $G_0$ not necessarily abelian), 
  together with an action $\rho_k$ of $G_0$ on $G_{k}$ for all $k \in \mathbb{N}$,
  such that 
  \begin{enumerate}
    \item 
	  $\rho_0 $ is the adjoint action of $G_0$ on itself;
	\item 
	  $\rho_1\circ \delta_0$ is the adjoint action of $G_1$ on itself;
	\item 
	  $\rho_k \circ \delta_0$ is the trivial action of $G_1$ on $G_k$
	  for $k > 1$;
	\item 
	  all $\delta_k$ respect the actions of $G_0$.
  \end{enumerate}
  A morphism of crossed complexes of groups
  $G_\bullet^\rho \to H_\bullet^{\rho'}$
  is a sequence of morphisms of component groups, respecting all this structure.
  Write $\mathrm{CrossedComplex}$ for the category of crossed complexes
  defined this way.
  \label{CrossedComplexOfGroups}
\end{definition}
\begin{remark}
If we write $\mathrm{ChainComplex}$ for the category of ordinary chain complexes
of abelian groups in non-negative degree, 
and $\mathrm{KanComplex} \hookrightarrow \mathrm{sSet}$
for the category of Kan complexes, then 
we have a diagram of (non-full) injections
$$
  \raisebox{20pt}{
  \xymatrix{
     \mathrm{ChainComplex} \ar[r] \ar[d]^\simeq 
     & 
	 \mathrm{CrossedComplex} 
	 \ar[r] \ar[d]^\simeq & 
	 \mathrm{KanComplex}
       \ar[d]^\simeq
     \\
     \mathrm{StrAb Str}\infty \mathrm{Grpd} 
     \ar[r] &
     \mathrm{Str} \infty \mathrm{Grpd}
     \ar[r] &
     \infty \mathrm{Grpd} \;\;,
  }
  }
$$
where in the top row we display models, and in the bottom row
the corresponding abstract notions. This immediately prolongs to
presheaves of complexes. Therefore every presheaf of 
crossed complexes $G_\bullet^\rho$ over $\mathrm{CartSp}$ presents a
smooth $\infty$-stack $\mathbf{B}(G_\bullet)$ in a way that restricts to the ordinary
Dold-Kan correspondence in the case that the crossed complex
is just an ordinary chain complex of abelian groups.
\end{remark}
\begin{examples}
  The $\mathrm{String}$ 2-group from 
  prop. \ref{BoundaryConfigsAreString2aConnections}
  has a presentation by the crossed complex 
  \footnote{See the appendix of \cite{FiorenzaSatiSchreiberII} for a more detailed review
of this and related results.}
  $$
    (G_1 \to G_0)
	:=
    (\hat \Omega \mathrm{Spin} \stackrel{\delta}{\to} P_* \mathrm{Spin})^\rho
	\,,
  $$
  where $P_* \mathrm{Spin}$ is the Fr{\'e}chet Lie group of based smooth paths in the 
	Lie group $\mathrm{Spin}$, where $\hat \Omega \mathrm{Spin}$
    is the Kac-Moody central extension of the group of smooth based
	loops in $\mathrm{Spin}$, where the morphism $\delta$ is the evident forgetful
	map, and where, finally, the action $\rho_1$ is given by a lift of the
	canonical conjugation action of paths on loops:
  $$
    \mathbf{B}\mathrm{String}
	  \simeq
	\mathbf{B}(\hat \Omega \mathrm{Spin} \to P_* \mathrm{Spin})^{\rho}
	\,.
  $$
\end{examples}
We are now interested in the following much simpler class of examples.
\begin{example}
  For every $n \in \mathbb{N}$, $n \geq 1$ 
  there is a crossed complex of groups of the form
  $$
    (U(1) \to 0 \to \cdots \to 0 \to \mathbb{Z}_2)^\rho
	\,,
  $$
  with $U(1)$ in degree $n$,  
  with all morphisms trivial and with $\mathbb{Z}_2$ acting in the canonical
  way on $U(1)$ via the identification
  $\mathrm{Aut}(U(1)) \simeq \mathbb{Z}_2$.
  With a slight abuse of notation we will write
  $$
    \mathbf{B}^{n+1} U(1)/\!/ \mathbb{Z}_2
	:=
	\mathbf{B}(U(1) \to 0 \to \cdots \to 0 \to \mathbb{Z}_2)^\rho
	\;\;
	\in \mathbf{H}
  $$
  for the corresponding moduli $(n+1)$-stacks.
\end{example}
\begin{remark}
  The 2-stack $\mathbf{B}^2 U(1)/\!/\mathbb{Z}_2$ is that 
  of \emph{$U(1)$-gerbes} in the original sense of Giraud.
  The canonical morphism of moduli 2-stacks
  $$
    \mathbf{B}^2 U(1) \to \mathbf{B}^2 U(1)/\!/\mathbb{Z}_2
  $$
  embeds the less general
  \emph{$U(1)$-bundle gerbes} into the genuine $U(1)$-gerbes.
This distinction is often not recognized in the literature,
  but in the following it makes all the difference. See also 
  observation \ref{JandlGerbes} further below.  
\end{remark}
\begin{definition}
  Define on $\mathbf{B}^n U(1)/\!/\mathbb{Z}_2$ the universal smooth
  characteristic map
  $$
    \mathbf{J}_{n-1} : \mathbf{B}^n U(1)/\!/\mathbb{Z}_2 \to \mathbf{B}\mathbb{Z}_2
  $$
  representing a universal class 
  $$
    [J] \in H^1(B^n U(1)/\!/\mathbb{Z}_2, \mathbb{Z}_2)
	\,,
  $$
  defined by the delooping of the evident morphism of crossed complexes
  $$
    (U(1) \to \cdots \to \mathbb{Z}_2)^\rho \to \mathbb{Z}_2
	\,.
  $$
\end{definition}
\begin{proposition}
  For all $n \geq 2$ there is a fiber sequence of smooth higher stacks
  $$
    \xymatrix{
      \mathbf{B}^n U(1) 
	  \ar[r] 
	  &
	  \mathbf{B}^n U(1)/\!/\mathbb{Z}_2 
	  \ar[r]^<<<<<{\mathbf{J}_{n-1}}
	  &
	  \mathbf{B}\mathbb{Z}_2
	  }
	\,.
  $$
\end{proposition}
\proof
  The canonical presentation of the morphism on the right by
  a morphism of simplicial presheaves is evidently a projective fibration.
  The claim then follows from the fact that $U(1)[n]$ is the fiber of
  the canonical morphism of crossed complexes
  $(U(1) \to \cdots \to \mathbb{Z}_2)^\rho \to (\cdots \to 0 \to \mathbb{Z}_2)$.
\endofproof
\begin{remark}
  This means that the morphism 
  $\mathbf{B}^n U(1) \to \mathbf{B}^n U(1)/\!/ \mathbb{Z}_2$
  exhibits a universal  double cover / universal $\mathbb{Z}_2$-principal bundle
  over the $n$-stack $\mathbf{B}^n U(1)/\!/ \mathbb{Z}_2$.
\end{remark}
\begin{corollary}
  \label{HigherOrientifoldsDecomposed}
  For $X \in \mathbf{H}$ any smooth space, a cocycle $g : X \to \mathbf{B}^n U(1)/\!/\mathbb{Z}_2$
  induces 
  \begin{enumerate}
   \item
     a choice of double cover $\hat X \to X$,
   \item
     a circle $n$-bundle $P$ over $\hat X$ equipped with a
	 $\mathbb{Z}_2$-twisted equivariance under the canonical $\mathbb{Z}_2$-action
	 on $\hat X$, such that
   \item 
      the restriction of $P$ to any fiber $\hat X_x$ of $\hat X$
	 is equivalent to the $n$-group extension
	 $(U(1) \to \cdots \to \mathbb{Z}_2)^\rho \to \mathbb{Z}_2$.
  \end{enumerate}
\end{corollary}
\proof
  Consider the induced pasting diagram of homotopy pullbacks, using
  prop. \ref{PastingLawForPullbacks}
  $$
    \raisebox{30pt}{
    \xymatrix{
	  (U(1) \to \cdots \to \mathbb{Z}_2)^\rho
	  \ar[r]
	  \ar[d]
	  &
	  P
	  \ar[r]
	  \ar[d]
	  &
	  {*}
	  \ar[d]
	  \\
	  \mathbb{Z}_2
	  \ar[r]
	  \ar[d]
	  &
	  \hat X
	  \ar[r]
	  \ar[d]
	  &
	  \mathbf{B}^n U(1)
	  \ar[r]
	  \ar[d]
	  &
	  {*}
	  \ar[d]
	  \\
	  {*}
	  \ar[r]^x
	  &
	  X \ar[r]^<<<<<g 
	  &
	  \mathbf{B}^n U(1)/\!/\mathbb{Z}_2
	  \ar[r]^<<<<{\mathbf{J}_{n-1}}
	  &
	  \mathbf{B}\mathbb{Z}_2
	}
	}
  $$
  \vspace{-5mm}
\endofproof

\begin{observation}
  For $n = 2$ the bundle gerbe incarnation of the 
  structures in corollary \ref{HigherOrientifoldsDecomposed}
  have been called \emph{Jandl bundle gerbes} \cite{SSW} and shown to encode 
  \emph{orientifold} backgrounds for strings.
  \label{JandlGerbes}
\end{observation}
We will discuss the analog for membranes. First we consider the
differential refinement of this situation. For this we need 
the following refinement of def. \ref{CrossedComplexOfGroups}.
\begin{definition}
  \label{CrossedComplex}
  A \emph{crossed complex of groupoids}  is a diagram
  $$
    C_\bullet
    =
    \left(
      \raisebox{20pt}{
      \xymatrix{
        \cdots
        \ar[r]^{\delta}
        &
        C_3
        \ar[r]^\delta
        \ar[d]
        &
        C_2
        \ar[r]^\delta
        \ar[d]
        &
        C_1
        \ar@<+3pt>[r]^{\delta_t}
        \ar@<-3pt>[r]_{\delta_s}
        \ar[d]^{\delta_s}
        &
        C_0
        \ar@{=}[d]
        \\
        \cdots
        \ar@{=}[r]
        &
        C_0
        \ar@{=}[r]
        &
        C_0
        \ar@{=}[r]
        &
        C_0
        \ar@{=}[r]
        &
        C_0
      }
      }
    \right),
  $$
  where
$\xymatrix{
  C_1
  \ar@<+3pt>[r]^{\delta_t}
  \ar@<-3pt>[r]_{\delta_s}
  &
  C_0
}$
is equipped with the structure of a groupoid, and where 
$C_k \to C_0$,
for all $k \geq 2$, are bundles of groups,
abelian for $k \geq 2$;
and equipped with a groupoid action $\rho_k$ of $C_1$, such that
\begin{enumerate}
\item
 the maps $\delta_k$, $k \geq 2$ are morphisms of groupoids over
 $C_0$ compatible with the action by $C_1$;
\item
  $\delta_{k-1} \circ \delta_k = 0$; k $\geq 3$;
\item
  $\rho_1$ is the conjugation action of the groupoid on its first homotopy groups;
\item 
  $\rho_2 \circ \delta_2 $ is the conjugation action of 
  $C_2$ on itself;
\item 
  $\rho_k \circ \delta_2 $ is the trivial action of $C_2$ for $k \geq 3$.
\end{enumerate}
A morphism of crossed complexes of groupoids is a sequence of morphisms
of the components such that all this structure is preserved.
\end{definition}
The nonabelian generalization of the Dold-Kan correspondence, 
reviewed in detail in \cite{BHS}, is now the following.
\begin{proposition}
  The category of crossed complexes of groupoids is equivalent to that 
  of strict globular $\infty$-groupoids.
  Under the natural simplicial nerve operation these embed into Kan complexes.
\end{proposition}
\begin{example}
  For $G_\bullet^\rho$ a crossed complex of groups, 
  def. \ref{CrossedComplexOfGroups}, we obtain a crossed complex
  of groupoids of the form
  $$
    \xymatrix{
	  \cdots
	  \ar[r]
	  &
	  G_2
	  \ar[r]
	  &
	  G_1
	  \ar[r]
	  &
	  G_0
	  \ar@<-4pt>[r]
	  \ar@<+4pt>[r]
	  &
	  {*}
	}
	\,.
  $$
  If $G_\bullet^\rho$ is a presheaf of crossed complexes of groupoids
  presenting a smooth $\infty$-group to be denoted by the same symbols, 
  then, under the nonabelian Dold-Kan correspondence this 
  presheaf of crossed complexes of groupoids presents the 
  smooth delooping $\infty$-stack $\mathbf{B}(G_\bullet^\rho)$.  
\end{example}
The example that we are interested here is the following.
\begin{definition}
  For $n \geq  2$ write
  $\mathbf{B}^n U(1)_{\mathrm{conn}}/\!/\mathbb{Z}_2$
  for the smooth $n$-stack presented by the presheaf of $n$-groupoids which 
  is given by the presheaf of crossed complexes of groupoids
  $$
    \xymatrix@C=3.3em{
      \Omega^n(-)\times C^\infty(-,U(1))
	  \ar[rr]^{\phantom{mm}(\mathrm{id},d_{\mathrm{dR}} \mathrm{log})}
	  &&
	  \Omega^n(-)\times \Omega^1(-)
	  \ar[r]^{\phantom{mmm}(\mathrm{id},d_{\mathrm{dR}})} &
	  \cdots 
	  \ar[r]^{(\mathrm{id},d_{\mathrm{dR}})\phantom{mmmmm}}
	  &
	  \Omega^n(-) \times \Omega^{n-2}(-)
	  \ar[r]^{\phantom{mmmmm}(\mathrm{id}, d_{\mathrm{dR}})}&
	  }
  $$
 $$
    \xymatrix@C=3.3em{
    \ar[r]^{(\mathrm{id}, d_{\mathrm{dR}})\phantom{mmmmmmmm}}
	  &
	  \Omega^n(-)\times \Omega^{n-1}(-) \times \mathbb{Z}_2
	  \ar@<-3pt>[r]
	  \ar@<+3pt>[r]
	  &
	  \Omega^n(-)
	}
	\,,
  $$
  where 
  \begin{enumerate}
    \item 
    the groupoid on the right has as morphisms $(A,\sigma) :  B \to B'$
	between two $n$-forms $B,B'$ pairs consisting of an $(n-1)$-form $A$
    and an element $\sigma \in \mathbb{Z}_2$, such that
	$(-1)^{\sigma} B' = B + d A$;
	\item
	  the bundles of groups on the left are all trivial as bundles;
	\item 
	  the $\Omega^1(-)\times\mathbb{Z}_2$-action is 
	  by the $\mathbb{Z}_2$-factor only and on forms given by multiplication 
	  by $\pm 1$ and on $U(1)$-valued functions by complex conjugation
	  (regarding $U(1)$ as the unit circle in the complex plane).
  \end{enumerate}
\end{definition}
\begin{observation}
  There are evident morphisms of smooth $n$-stacks
  $$
    \mathbf{\flat} \mathbf{B}^n U(1)/\!/\mathbb{Z}_2
	\to
	\mathbf{B}^n U(1)_{\mathrm{conn}}/\!/\mathbb{Z}_2
	\to
	\mathbf{B}^n U(1)/\!/\mathbb{Z}_2
	\,,
  $$
  where the first one includes the flat differential coefficients 
  and the second one forgets the connection.
\end{observation}
\begin{remark}
  A detailed discussion of $\mathbf{B}^2 U(1)_{\mathrm{conn}}/\!/\mathbb{Z}_2$
  is in \cite{SWII} and \cite{SWIII}. 
\end{remark}
We now discuss differential cocycles with coefficients in 
$\mathbf{B}^n U(1)_{\mathrm{conn}}/\!/\mathbb{Z}_2$ 
over $\mathbb{Z}_2$-quotient stacks / orbifolds. Let $Y$ be a smooth manifold 
equipped with a smoth $\mathbb{Z}_2$-action $\rho$. Write 
  $Y /\!/ \mathbb{Z}_2$ for the corresponding global orbifold and 
  $\rho : Y/\!/\mathbb{Z}_2 \to \mathbf{B}\mathbb{Z}_2$
  for its classifying morphism, hence for the morphism that fits into 
  a fiber sequence of smooth stacks
  $$
    \xymatrix{
	  Y \ar[r] 
	  & 
	  Y/\!/\mathbb{Z}_2
	  \ar[r]
	  &
	  \mathbf{B}\mathbb{Z}_2
	}
	\,.
  $$
\begin{definition}
   An \emph{$n$-orientifold} structure $\hat G_{\rho}$ on 
   $(Y, \rho)$ is a \emph{$\rho$-twisted 
  $\hat {\mathbf{J}}_n$-structure}
  on $Y/\!/\mathbb{Z}_2$, def. \ref{TwistedCStructures},
  hence a dashed morphism in the diagram
  $$  
    \xymatrix{
	  & \mathbf{B}^{n+1} U(1)_{\mathrm{conn}}/\!/\mathbb{Z}_2
	  \ar[d]^{\hat {\mathbf{J}}_n}
	  \\
	  Y/\!/\mathbb{Z}_2
	  \ar@{-->}[ur]^{\hat G_{\rho}}
	  \ar[r]^{\rho}
	  &
	  \mathbf{B}\mathbb{Z}_2
	}
	\,.
  $$
\end{definition}
\begin{observation}
  By corollary \ref{HigherOrientifoldsDecomposed},
  an $n$-orientifold structure decomposes into 
  an ordinary $(n+1)$-form connection $\hat G$ on a circle $(n+1)$-bundle over
  $Y$, subject to a $\mathbb{Z}_2$-twisted $\mathbb{Z}_2$-equivariance
  condition
  $$
    \raisebox{20pt}{
    \xymatrix{
	  Y
	  \ar[r]^<<<<<{\hat G}
	  \ar[d]
	  &
	  \mathbf{B}^{n+1} U(1)_{\mathrm{conn}}
	  \ar[r]
	  \ar[d]
	  &
	  {*}
	  \ar[d]
	  \\
	  Y/\!/\mathbb{Z}_2
	  \ar[r]^<<<{\hat G_\rho}
	  \ar@/_1pc/[rr]_{\rho}
	  &
	  \mathbf{B}^{n+1} U(1)_{\mathrm{conn}}/\!/\mathbb{Z}_2
	  \ar[r]^<<<<<{\hat {\mathbf{J}_n}}
	  &
	  \mathbf{B}\mathbb{Z}_2\;\;.
	}
	}
  $$
  For $n = 1$ this reproduces, via observation \ref{JandlGerbes},
  the \emph{Jandl gerbes with connection} from \cite{SSW}, hence
  ordinary string orientifold backgrounds, as discussed there.
  \label{DifferentialJStructreDecomposed}
\end{observation}
\begin{observation}
  Let $U/\!/\mathbb{Z}_2 \hookrightarrow Y/\!/\mathbb{Z}_2$
  be a patch on which a given $\hat {\mathbf{J}}_n$-structure 
  has a trivial underlying integral class, such that it is
  equivalent to a globally defined $(n+1)$-form $C_U$ on $U$. Then 
  the components of this 
  this 3-form orthogonal to the $\mathbb{Z}_2$-action are 
  \emph{odd} under the action.
  In particular, if $U \hookrightarrow Y$ sits in the fixed point
  set of the action, then these components vanish. 
  This is the Ho{\v r}ava-Witten boundary condition on the $C$-field
  on an 11-dimensional spacetime $Y = X \times S^1$ equipped with 
  $\mathbb{Z}_2$-action on the circle. See for instance section 3
  of \cite{Falkowski} for an explicit discussion of the $\mathbb{Z}_2$
  action on the C-field in this context.
\end{observation}
We therefore have a natural construction of the moduli 3-stack of 
Ho{\v r}ava-Witten C-field configurations as follows
\begin{definition}
 Let $\mathbf{CField}_J(Y)$ be the homotopy pullback in 
 $$
   \raisebox{20pt}{
  \xymatrix{
    \mathbf{CField}_J(Y)
    \ar[dd]
	\ar[rr]
    &&
    \hat {\mathbf{J}}_2 \mathrm{Struc}_{\rho}(Y/\!/\mathbb{Z}_2)
	\ar[d]
	\\
	&& \mathbf{H}(Y, \mathbf{B}^3 U(1)_{\mathrm{conn}})
	\ar[d]
	\\
	\mathbf{H}(Y, \mathbf{B}\mathrm{Spin}_{\mathrm{conn}} \times \mathbf{B}E_8)
	 \ar[rr]^<<<<<<<<<<{\mathbf{H}(Y, \tfrac{1}{2}\mathbf{p}_1 + 2\mathbf{a})} 
	&& \mathbf{H}(Y, \mathbf{B}^3 U(1))\;\;,
  }
  }
 $$
 where the top right morphism is the map $\hat G_\rho \mapsto \hat G$
 from remark \ref{DifferentialJStructreDecomposed}.
\end{definition}
The objects of $\mathbf{CField}_J(Y)$ are C-field configurations
on $Y$ that not only satisfy the flux quantization condition, but also
the Ho{\v r}ava-Witten twisted equivariance condition
(in fact the proper globalization of that condition from 3-forms to 
full differential cocycles). This is formalized by the following.
\begin{observation}
  There is a canonical morphism $\mathbf{CField}_J(Y) \to \mathbf{CField}(Y)$,
  being the dashed morphism in
 $$
   \raisebox{20pt}{
  \xymatrix{
    \mathbf{CField}_J(Y)
    \ar@{-->}[d]
	\ar[rr]
    &&
    \hat {\mathbf{J}}_2 \mathrm{Struc}_{\rho}(Y/\!/\mathbb{Z}_2)
	\ar[d]
	\\
	\mathbf{CField}(Y) 
	  \ar[rr]
	  \ar[d]
	&& \mathbf{H}(Y, \mathbf{B}^3 U(1)_{\mathrm{conn}})
	\ar[d]
	\\
	\mathbf{H}(Y, \mathbf{B}\mathrm{Spin}_{\mathrm{conn}} \times \mathbf{B}E_8)
	 \ar[rr]^<<<<<<{~~~~~~~\mathbf{H}(Y, \tfrac{1}{2}\mathbf{p}_1 + 2\mathbf{a})} 
	&& \mathbf{H}(Y, \mathbf{B}^3 U(1))\;\;,
  }
  }
  \,
 $$
  which is given by the universal property of the defining homotopy pullback
  of $\mathbf{CField}$, remark \ref{CFieldAs3Bundle}.
\end{observation}
A supergravity field configuration presented by a morphism $Y \to \mathbf{CField}$
into the moduli 3-stack of configurations that satisfy the flux quantization
condition in addition satisfies the Ho{\v r}ava-Witten boundary condition
if, as an element of $\mathbf{CField}(Y) := \mathbf{H}(Y, \mathbf{CField})$
it is in the image of $\mathbf{CField}_J(Y) \to \mathbf{CField}(Y)$.
In fact, there may be several such pre-images. A choice of one is a choice
of membrane orientifold structure.

\vspace{5mm}
\noindent
%%%%%%%%%%%%%%%%%%%%%%%%%%%%%%%%%%%%%%%%%%%%%%%%%%
{\bf \large Acknowledgements}
%%%%%%%%%%%%%%%%%%%%%%%%%%%%%%%%%%%%%%%%%%%%%%%%%%

\vspace{1mm}

\noindent The research of H.S. is supported by NSF Grant PHY-1102218.
 U.S. gratefully acknowledges an invitation 
 to the Department of Mathematics of Pittsburgh University
 in September 2011, which helped the work presented here
 come into existence.

%%%%%%%%%%%%%%%%%%%%%%%%%%%%%%%%%%%%%%%%%%%%%%%%%%%

\end{document}